\documentclass[lettersize,journal]{IEEEtran}
\usepackage{amsmath,amsfonts}
\usepackage{algorithmic}
\usepackage{algorithm}
\usepackage{array}
\usepackage{textcomp}
\usepackage{stfloats}
\usepackage{url}
\usepackage{verbatim}
\usepackage{graphicx}
\usepackage{cite}
\usepackage{amssymb}

\usepackage{xcolor}

\usepackage{booktabs}
\usepackage{makecell}
\usepackage{multirow}
\usepackage{adjustbox}
\usepackage{threeparttable}
\usepackage{arydshln}

\usepackage{float}
\usepackage{subcaption}

\usepackage{pifont}
\usepackage{soul}
\usepackage{ragged2e}

\usepackage{enumitem}

\usepackage{url}

\begin{document}

\title{BioMoTouch: Touch-Based Behavioral Authentication via Biometric-Motion Interaction Modeling}

\author{
Zijian~Ling,
Jianbang~Chen,
Hongwei~Li,
Hongda~Zhai,
Man~Zhou,
Jun~Feng,
Zhengxiong~Li,\\
Qi~Li,~\IEEEmembership{Senior Member,~IEEE,}
and~Qian~Wang,~\IEEEmembership{Fellow,~IEEE}

%\thanks{Z. Ling and J. Chen contributed equally to this work. M. Zhou's work is supported by the NSFC under Grant .  (\emph{Corresponding author: Man Zhou})}

\thanks{Z. Ling, J. Chen, H. Li, H. Zhai, M. Zhou, and J. Feng are with the
Hubei Key Laboratory of Distributed System Security,
Hubei Engineering Research Center on Big Data Security,
School of Cyber Science and Engineering,
Huazhong University of Science and Technology,
Wuhan 430074, China
(e-mail: \{zijianling, jianbangchen, hongweili, zhd, zhouman, junfeng\}@hust.edu.cn).}
\thanks{ Z. Li is with the Department of Computer Science and Engineering, University of Colorado Denver, Denver, CO 80204 USA (e-mail: zhengxiong. li@ucdenver.edu).}
\thanks{ Q. Li is with the Institute for Network Sciences and Cyberspace, Tsinghua University, Beijing 100084, China (E-mail: qli01@tsinghua.edu.cn).}
\thanks{Q. Wang is with the Key Laboratory of Aerospace Information Security and Trusted Computing, Ministry of Education, School of Cyber Science and
Engineering, Wuhan University, Wuhan 430072, China (e-mail: qianwang@whu.edu.cn).}

}

\IEEEtitleabstractindextext{%

\maketitle

\begin{abstract}
Touch-based authentication is widely deployed on mobile devices due to its convenience and seamless user experience. However, existing systems largely model touch interaction as a purely behavioral signal, overlooking its intrinsic multi-dimensional nature and limiting robustness against sophisticated adversarial behaviors and real-world variations.
In this work, we present BioMoTouch, a multi-modal touch authentication framework on mobile devices grounded in a key empirical finding: during touch interaction, inertial sensors capture user-specific behavioral dynamics, while capacitive screens simultaneously capture physiological characteristics related to finger morphology and skeletal structure. 
Building upon this insight, BioMoTouch jointly models physiological contact structures and behavioral motion dynamics by integrating capacitive touchscreen signals with inertial measurements. Rather than combining independent decisions, the framework explicitly learns their coordinated interaction to form a unified representation of touch behavior.
BioMoTouch operates implicitly during natural user interactions and requires no additional hardware, enabling practical deployment on commodity mobile devices. We evaluate BioMoTouch with 38 participants under realistic usage conditions. Experimental results show that BioMoTouch achieves a balanced accuracy of 99.71\% and an equal error rate of 0.27\%. Moreover, it maintains false acceptance rates below 0.90\% under artificial replication, mimicry, and puppet attack scenarios, demonstrating strong robustness against partial-factor manipulation.

\end{abstract}

}

\maketitle
\IEEEdisplaynontitleabstractindextext
\IEEEpeerreviewmaketitle

%TODO: 修改思路，前面的background不变，但是静态生物特征的说法要修改掉。说完触控交互的优点后，“之前的研究都是认为，触控交互仅包括行为，我们发现触控交互既包括生理也包括行为。”这点我觉得可以放在challenge和技术创新上
%因此就需要改下challenge 1和2，比如将challenge1改为，对触控交互的多维度、深层次特征提取，然后在这里面说

%TODO：前面的背景介绍有些多余
\section{Introduction}

With the widespread adoption of smart mobile devices, biometric authentication has become a core component of modern mobile security systems. Among existing authentication approaches~\cite{lien2023iotbiometrics}, fingerprint and face recognition remain the most widely deployed techniques and are commonly integrated into consumer devices by major technology vendors, such as Apple’s Touch ID~\cite{apple_touch_id} and Face ID~\cite{apple_faceid}, and Samsung’s in-display ultrasonic fingerprint sensing solutions~\cite{samsung_ultrasonic_fp}.
However, most mainstream biometric authentication methods fundamentally rely on static biometric traits, such as fingerprint textures~\cite{jia2024finger} or facial features~\cite{yan2022age}. Beyond the privacy concerns they raise~\cite{techreview2019biometricleak}, the static nature of these traits also exposes inherent security risks~\cite{zhou2026stealing}. In particular, artificial replication attacks—where adversaries fabricate physical replicas of fingerprints or facial features—have repeatedly demonstrated the feasibility of bypassing such systems~\cite{casula2024realistic,Li_2023_CVPR}. 
Once biometric information is exposed, the resulting threat is permanent, as biometric traits cannot be revoked or replaced. Consequently, authentication systems that rely on static biometrics inherently suffer from long-term security risks.

To mitigate these limitations, recent studies have introduced liveness detection mechanisms, such as sensing subsurface structures (e.g., finger~\cite{huang2023fvfsnet}, palm veins~\cite{li2025palmvein}), blood flow~\cite{Kossack_2022_CVPR}, or other physiological signals beneath the skin~\cite{zhu2025caphandauth}. 
While such approaches can effectively defend against basic spoofing attacks, they typically rely on specialized hardware, including infrared cameras~\cite{li2025palmvein} or large capacitive touchscreens~\cite{zhu2025caphandauth}. This hardware dependency substantially limits their deployability on commodity devices and often incurs additional authentication latency, imposing extra user burden~\cite{huang2022pcrauth}.
The broader collection of static biometric information further increases the risk of privacy exposure and biometric information leakage.
More critically, puppet attacks~\cite{251562} represent a potent threat model, in which the attacker can forcibly manipulate an unaware user into performing biometric authentication.
Under this threat model, even advanced liveness detection systems can be bypassed, exposing a fundamental weakness shared by existing biometric defenses that rely on static biometric traits.

In contrast, touch interaction arises naturally during device usage and provides rich interaction-level signals beyond static biometric inputs~\cite{olugbade2023touchsurvey,wu2024host}. During capacitive touchscreen contact, sensing responses are influenced by multiple factors, including contact structure, finger-surface interaction patterns, and hand motion dynamics. These factors jointly shape observable touch signals in complex and subtle ways.
However, existing studies have not fully exploited this potential. Most prior work models touch interaction from a single perspective (e.g., behavioral dynamics or contact-induced physiological responses), without capturing their coordinated relationship.

Motivated by this gap, we revisit touch interaction as an intrinsically multi-dimensional authentication signal and explicitly model the interaction between its physiological and behavioral components using commodity mobile sensors. Our goal is to achieve hardware-free, implicit, and robust authentication in natural usage scenarios. Achieving this goal raises two core challenges:

%TODO：可能需要精简
\textbf{Challenge 1: Multi-dimensional and deep modeling of touch interaction.}
Most existing touch-based authentication methods treat touch interaction primarily as a behavioral signal, focusing on dynamic interaction features (e.g., Fingerbeat~\cite{10443592}, PressPIN~\cite{9714878}). However, capacitive touchscreen signals are jointly shaped by intrinsic physiological constraints (e.g., finger morphology and contact structure) and dynamic motor behaviors.
Although some studies attempt to capture physiological characteristics via built-in vibration motors~\cite{xu2020touchpass,10251599}, they often emphasize structural signals while overlooking concurrent behavioral dynamics. Moreover, vibration-based approaches are sensitive to environmental disturbances (e.g., screen protectors, device casing, and mechanical vibrations), limiting their robustness in practice.
Therefore, effectively capturing and modeling the intrinsic coupling between physiological structure and behavioral dynamics using commodity sensors remains a fundamental challenge.

\textbf{Challenge 2: Preserving discriminative robustness under partial-factor manipulation.}
% In realistic adversarial scenarios, attackers may compromise only one dimension of touch interaction. For instance, artificial replication attacks attempt to reproduce physiological traits, mimicry attacks focus on imitating behavioral patterns, and puppet attacks manipulate genuine physiological inputs through externally induced actions.
% A straightforward solution might be to independently model physiological and behavioral components and combine their decisions at the authentication stage (e.g., via logical conjunction). However, such decision-level aggregation implicitly assumes independence between the two dimensions and does not explicitly capture their intrinsic coupling during natural touch interaction.
% When one component is partially controlled or manipulated, independently modeled systems may still be vulnerable to crafted interactions that appear plausible in each dimension but deviate from their natural joint distribution. 
% Therefore, designing a mechanism that explicitly models and preserves the intrinsic physiological-behavioral coupling—rather than merely combining independent decisions—constitutes a critical security-level challenge.
In realistic adversarial scenarios, attackers may compromise only one dimension of touch interaction, such as replicating physiological traits, mimicking behavioral patterns, or manipulating genuine inputs via external actuation. 
A straightforward solution is to model physiological and behavioral components independently and combine their decisions (e.g., via logical conjunction~\cite{10443592}). However, such decision-level aggregation implicitly assumes independence and fails to capture their intrinsic coupling during natural interaction. 
As a result, inputs may appear valid in each modality individually while remaining inconsistent when considered jointly, allowing such attacks to bypass independent decision rules. 
Therefore, explicitly modeling and preserving physiological-behavioral coupling is essential for robust authentication.

To address the above challenges, we propose BioMoTouch, a multi-modal touch-based authentication framework that models the coupling between physiological contact structures and behavioral motion dynamics. This design is grounded in our empirical finding that finger morphology and skeletal structure induce distinctive physiological patterns in capacitive responses. Our evaluation further shows that such features alone are discriminative for identity authentication.
%Capacitive data reflect physiological contact structures, while IMU signals encode motion dynamics, and their coordinated interaction is learned to form a unified representation.
At the sensing stage, BioMoTouch introduces a lightweight, training-free touch-localization and tracking scheme that robustly isolates genuine contact from noisy capacitive streams, addressing the challenge of reliable signal acquisition under real-world disturbances. To preserve cross-modal correspondence, we further propose a coarse-to-fine temporal refinement strategy that aligns capacitive events with their corresponding IMU responses, enabling consistent multi-dimensional modeling of touch interactions. Based on the refined segments, physiological and behavioral representations are extracted, with IMU signals enhanced via quaternion-based motion estimation and time–frequency analysis.
At the representation-learning stage, modality-specific feature extractors are employed to learn high-level embeddings, capturing complementary physiological and behavioral characteristics. To improve robustness under partial-factor manipulation, we introduce capacitive-side augmentation to simulate realistic variations while preserving touch structure. The learned features are integrated through an interaction-aware fusion module that explicitly models their coordinated relationship, overcoming the limitations of independent modeling. %The resulting representation is fed into a user-specific one-class classifier.

We conducted a comprehensive evaluation of BioMoTouch through a large-scale user study involving 38 participants under natural interaction settings. The experimental results show that BioMoTouch achieves strong authentication performance, reaching a balanced accuracy (BAC) of 99.71\% and an equal error rate (EER) of 0.27\% under the default setting.
BioMoTouch also demonstrates robust resistance against advanced adversarial behaviors. Across artificial replication, mimicry, and puppet attack scenarios, it consistently maintains false acceptance rates below 0.90\%, substantially outperforming existing solutions.
In addition, BioMoTouch can be designed as an auxiliary behavioral signal, seamlessly integrated into existing touch-involved authentication workflows. Our evaluation shows that BioMoTouch remains effective when paired with primary authentication mechanisms with touch action, including fingerprint- and PIN-based unlocking, achieving EERs of 0.27\% and 0.19\%, respectively. %This indicates that touch behaviors captured by BioMoTouch are highly consistent across different authentication contexts, enabling practical and deployment-friendly integration without modifying existing authentication pipelines.

This work makes the following key contributions:
\begin{itemize}
    \item  
    We propose BioMoTouch, a multi-modal touch-based authentication framework that jointly leverages capacitive touchscreen data and inertial measurements on mobile devices to capture rich touch interaction behaviors. The proposed design requires no additional hardware and operates implicitly during natural user interactions, making it readily deployable on commodity mobile devices.

    \item 
    To our knowledge, we are the first to empirically demonstrate that commodity capacitive screens can directly capture user-specific physiological characteristics arising from finger morphology and skeletal structure during natural touch interaction. This finding motivates a multi-dimensional modeling approach that explicitly captures the coordinated interaction between physiological and behavioral signals.

    \item
    We conduct a large-scale evaluation involving 38 participants under diverse real-world conditions. BioMoTouch achieves strong authentication performance, with a balanced accuracy (BAC) of 99.71\% and an equal error rate (EER) of 0.27\%.
    The results validate the robustness, stability, and practicality of BioMoTouch in realistic deployment scenarios. 

    \item
    We systematically demonstrate that modeling multi-dimensional touch interaction dynamics substantially improves robustness against advanced adversarial behaviors, including mimicry attack, artificial replication attack, and the most challenging puppet attack. Across all evaluated attack scenarios, BioMoTouch consistently maintains false acceptance rates below 0.90\%.

    %We design BioMoTouch as a flexible behavioral authentication mechanism that can operate both as a standalone authenticator and as an auxiliary signal integrated into existing touch-involved authentication workflows. Extensive experiments show that BioMoTouch remains effective when combined with different primary authentication mechanisms, including fingerprint- and PIN-based unlocking, achieving EERs of 0.27\% and 0.19\%, respectively, without requiring any modification to existing authentication pipelines.

   % \item \textbf{Comprehensive Real-World Evaluation.} 
    %We conduct a large-scale evaluation involving 38 participants under diverse real-world conditions. The results validate the robustness, stability, and practicality of BioMoTouch in realistic deployment scenarios.
\end{itemize}
%TODO：修改为一整个段落？尝试分为两个subsection

\section{Related Work}
\textbf{Behavioral Modeling of Touch Interaction.} 
Traditionally, touch interaction has been regarded as a form of behavioral biometric, where prior studies primarily focus on dynamic interaction patterns, including touch pressure~\cite{9714878}, contact location~\cite{wu2024host}, contact area~\cite{10.1145/3300061.3345434}, and motion or multimodal features~\cite{8006292,9737094}. Under this perspective, touch is treated mainly as a manifestation of users’ motor behavior during device interaction.

For instance, PressPIN~\cite{9714878} models the relationship between applied pressure and structure-borne sound degradation, deriving pressure values from PIN inputs to construct an $n$-digit pressure code for authentication. Wu et al.~\cite{wu2024host} analyze pressure and pressing location to explore behavioral characteristics for identity verification. Li et al.~\cite{10.1145/3300061.3345434} use the temporal variation of finger contact area during touch interactions for authentication. Shen et al.~\cite{8006292} characterize motion sensor signals using statistical, frequency-domain, and wavelet-domain features to represent user touch actions and evaluate their discriminability and stability across scenarios. MMAuth~\cite{9737094} integrates heterogeneous multimodal identity information, including motion patterns and touch dynamics, and proposes a time-extended behavioral feature set to improve authentication accuracy.
These representative studies collectively demonstrate the potential of behavioral modeling for touch-based authentication. However, the extracted interaction features are often confined to specific signal modalities or limited behavioral dimensions. Consequently, such approaches may face challenges in maintaining temporal consistency and robustness under varying environmental conditions, and their resilience against more sophisticated adversarial attacks remains constrained.

\textbf{Physiological Dimensions of Touch Interaction.}
Touch interaction is inherently a composite process that reflects both physiological characteristics and behavioral dynamics. When a user contacts a touchscreen, the resulting sensing signals are shaped not only by motor patterns but also by intrinsic biological traits. These include finger geometry~\cite{7958587,chen2020listentofingers}, skin surface properties~\cite{xu2020touchpass,10251599}, and physiological signals~\cite{9920171}, which together contribute to a richer and more holistic representation of user identity.

Several studies have investigated these intrinsic, relatively stable physiological features that manifest during touch. Beyond conventional fingerprint patterns, researchers have explored richer sensing mechanisms to capture such traits. Song et al.~\cite{7958587} associate multi-touch traces with finger geometry for user differentiation. TouchPrint~\cite{chen2020listentofingers} employs active acoustic sensing to infer finger geometry during pressing, while TouchPass~\cite{xu2020touchpass} and FingerSlid~\cite{10251599} utilize active vibration signals to characterize finger-surface contact properties. Wu et al.~\cite{9920171} further extract subtle vascular pulsation signals from static fingertip pressure for authentication. 
% Song et al.~\cite{7958587} infer hand structural features from touch interaction trajectories for user authentication, whereas Chen et al~\cite{chen2020listentofingers}. employ active acoustic sensing to estimate hand geometric characteristics for authentication.
These studies demonstrate that physiological traits can indeed be leveraged during touch interaction. However, many of them rely on additional hardware components or active sensing modules, which may increase deployment complexity and interaction overhead. Furthermore, physiological and behavioral signals are often modeled independently, without fully exploiting their natural coupling in real-world touch scenarios.

\textbf{Our Insight.}
In contrast, our empirical investigation reveals that during natural touchscreen contact, users’ unique finger morphology and skeletal structure give rise to distinctive capacitive sensing patterns that can be captured directly by commodity capacitive screens. As validated in our evaluation, capacitive sensing alone already exhibits discriminative power for identity recognition, confirming the feasibility of leveraging such signals in practice.
Building on this insight, BioMoTouch extracts physiological features from capacitive-sensing signals and behavioral features from IMU measurements, integrating them within a unified fusion framework on mobile devices. By jointly modeling these complementary dimensions, the proposed approach provides a more comprehensive representation of touch interaction and achieves state-of-the-art authentication performance.

\section{Threat Model}
In this work, we consider scenarios in which an attacker seeks to impersonate a legitimate user in order to bypass the authentication system and gain unauthorized access to sensitive information. We focus on several practical and representative attack scenarios that pose significant threats to the integrity and confidentiality of the authentication system.

\textbf{Mimicry Attack.}
In behavioral authentication systems, mimicry attack represents a practical and unavoidable threat model that must be considered. In such attacks, the adversary goes beyond merely observing authentication inputs and attempts to imitate the victim’s touch interaction behavior during authentication. Specifically, the attacker uses their own finger to bypass the system by mimicking the victim’s touch location, pressing pattern, and interaction rhythm.
In practice, such imitation can be guided by observing the victim’s authentication process, for example, through video recordings captured without the victim’s awareness in public settings. By iteratively observing, comparing, and adjusting their actions, the attacker can gradually approximate the behavioral characteristics of the legitimate user’s touch interactions, thereby increasing the likelihood of successful impersonation.

\textbf{Artificial Replication Attack.}
Artificial replication attack represents a more advanced threat model, in which the adversary seeks to impersonate a legitimate user by fabricating a physical replica of the victim’s biometric traits. 
In this attack, the adversary first acquires the victim’s target finger through means such as physical theft, unauthorized collection, or social engineering. The attacker then uses biometric imitation materials to construct a highly realistic artificial replica that closely reproduces the physical characteristics of the original finger. This fabricated replica is subsequently used to deceive the biometric authentication system, allowing the adversary to impersonate the legitimate user and gain unauthorized access.

\textbf{Puppet Attack.}
Puppet attack is a threat model that has been proposed in recent years and reported by multiple studies~\cite{251562,zhu2025caphandauth}, reflecting a more realistic and practical attack scenario. 
In this attack, the adversary forcibly employs the victim’s legitimate finger to perform the touch action without the victim’s consent, for instance, when the victim is asleep or unconscious. Compared with traditional spoofing attempts, a puppet attack represents a more advanced threat, as it directly exploits the victim’s genuine biometric traits. Consequently, existing liveness detection techniques are rendered ineffective in defending against such attacks~\cite{251562}.

% \textbf{Out-of-Scope.}
% From a practical standpoint, the following types of attacks are excluded from our consideration, as they are deemed nearly infeasible in real-world scenarios.
% Robot-based mimicry attack.
% An attacker may attempt to launch a sophisticated mimicry attack using a robotic system. 
% However, the unique behavioral biometric traits embedded in a user's touch interactions are inaccessible to attackers. This makes it prohibitively difficult to train a robotic system to mimic these traits with sufficient fidelity. In addition, current-generation robots are high-cost to \$14,990~\cite{robotlab2022nao}, further limiting the practicality of such attacks. Given that these issues are unlikely to be resolved in the near term, robot-based attack methods are excluded from the scope of this study.
\section{Method}

\begin{figure*}[t]
  \centering
  \includegraphics[width=0.95\linewidth]{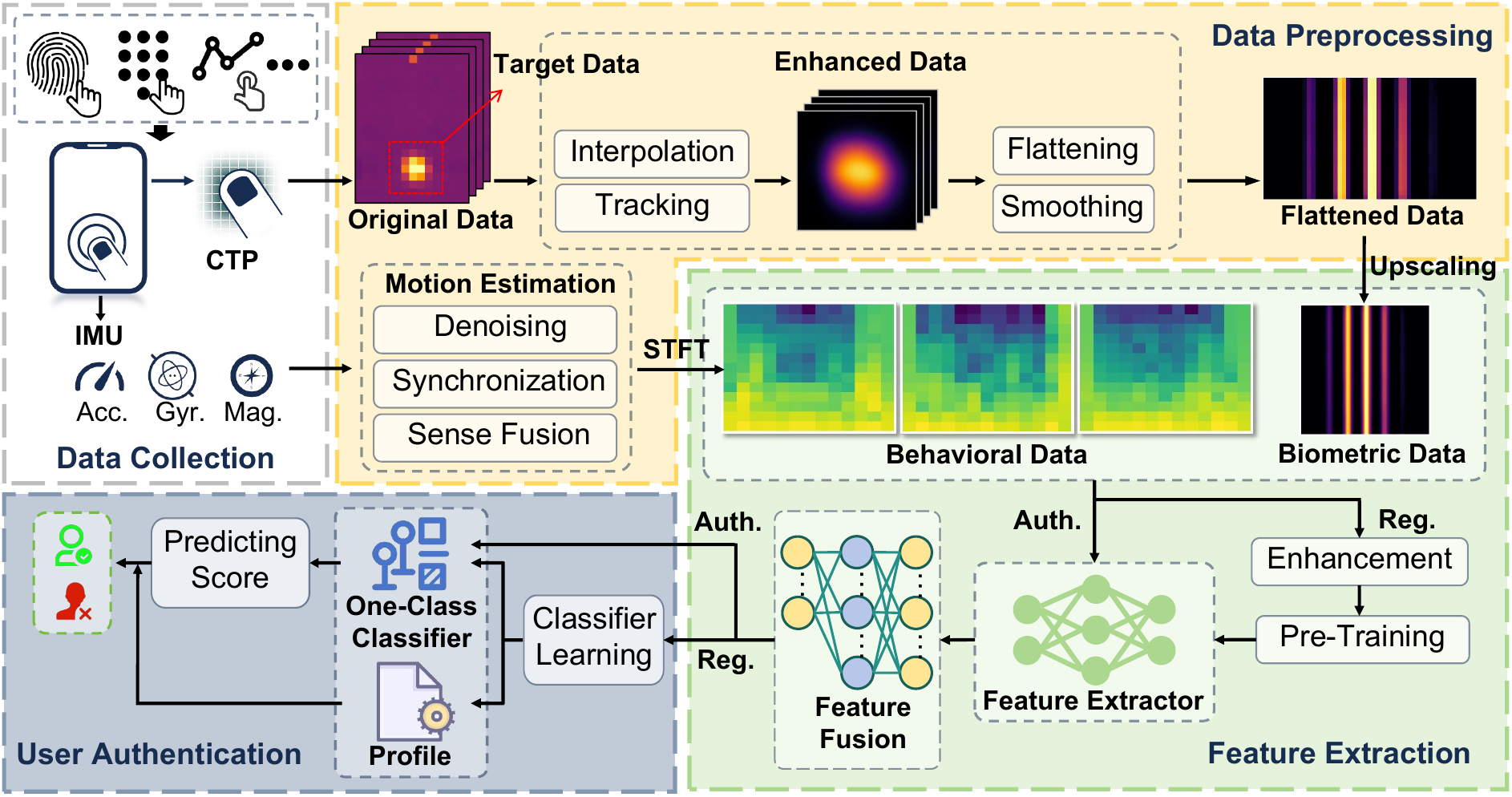} 
  \caption{The workflow of BioMoTouch.}
  \label{fig:pattern} 
\end{figure*}

\subsection{System Overview}
Our objective is to design an implicit, hardware-free touch authentication system that operates transparently during natural user interactions while remaining robust against advanced attack scenarios, including mimicry, artificial replication, and puppet attacks. Achieving this goal requires addressing two key challenges: (i) modeling the multi-dimensional nature of touch interaction beyond a purely behavioral perspective, and (ii) preserving discriminative robustness when either physiological or behavioral components are partially manipulated.

Through empirical investigation, we demonstrate that touch interactions on mobile devices involve two complementary sensing modalities: inertial sensors capture user-specific behavioral dynamics, while capacitive screens simultaneously capture physiological characteristics related to finger morphology and skeletal structure.
 %As validated in our evaluation, capacitive sensing alone already exhibits discriminative capability for identity recognition, confirming the feasibility of exploiting such structural signals in practice. 
Building upon this observation, BioMoTouch adopts a unified multimodal design that combines capacitive touchscreen signals with inertial measurements. Capacitive data reflect contact structures shaped by intrinsic physiological constraints, while IMU signals capture motion dynamics induced during touch operations. Rather than modeling these two dimensions independently, BioMoTouch explicitly learns their coordinated interaction to form a unified representation of touch behavior.

As shown in Fig.~\ref{fig:pattern}, BioMoTouch follows a unified end-to-end pipeline. During natural device usage, the system passively collects capacitive and IMU data without requiring additional user actions. A lightweight \emph{Data Preprocessing} stage first localizes genuine touch events and suppresses non-contact disturbances. The processed signals are then transformed into structured modality-specific representations that preserve spatial contact patterns, temporal evolution, and motion dynamics.
At the \emph{Feature Extraction} stage, BioMoTouch applies separate feature extractors to learn high-level representations from each modality. Instead of performing decision-level aggregation, the learned features are integrated through a lightweight fusion module that captures their coordinated structure and produces a unified embedding. The \emph{User Authentication} stage is formulated as a user-specific one-class classification problem, aligning with practical deployment settings where only legitimate user data is available during enrollment.
By explicitly modeling the intrinsic coupling between physiological structure and behavioral dynamics, BioMoTouch enhances robustness under partial-factor manipulation and improves resilience against advanced adversarial scenarios.

\subsection{Data Collection}
In the daily usage of mobile devices (e.g., smartphones and tablets), users frequently perform touch interactions during tasks such as fingerprint authentication, PIN entry, and routine operations. 
In these scenarios, BioMoTouch implicitly and unobtrusively captures the preliminary feature data of users’ touch behaviors, without imposing additional action burdens on the user or incurring extra hardware costs. 
Specifically, BioMoTouch leverages the device’s Capacitive Touch Panel (CTP) to extract the user’s finger physiological characteristics, while simultaneously utilizing the Inertial Measurement Unit (IMU)—comprising the accelerometer, gyroscope, and magnetometer—to comprehensively record the pressing behavior characteristics associated with touch interactions.
The preliminary touch behavior data are subsequently input into the Data Preprocessing module, where they are subjected to more fine-grained refinement for subsequent analysis.

\subsection{Data Preprocessing}
% TODO：这里加图，体现我们是怎么去除 water / sleeve 等的干扰，以及如何追踪定位算法

\subsubsection{Touch Detection}

To standardize the data representation, we first interpolate capacitive measurements to a fixed number of frames. In practice, touch behaviors are easily affected by various non-touch interferences. Our goal is to avoid falsely identifying variations caused by non-touch events (e.g., sliding water droplets, sleeve or clothing friction), while not relying on prior knowledge such as user-specific touch samples for training.
To achieve this, we adopt a lightweight, training-free method for localizing and tracking touch behaviors in capacitive frames.

Let $X_t \in \mathbb{R}^{H \times W}$ denote the capacitive frame at time index $t$, where
$X_t(i,j)$ represents the measurement at row $i$ and column $j$.
To suppress background fluctuations and non-touch interferences, we apply an adaptive
thresholding scheme based on robust statistics:
\begin{equation}
\tau_t = \operatorname{median}(X_t) + k \cdot \operatorname{MAD}(X_t),
\end{equation}
where $k>0$ controls the detection sensitivity and $\operatorname{MAD}(\cdot)$ denotes the
median absolute deviation.
Measurements exceeding $\tau_t$ are preserved as touch-related responses, while low-amplitude background variations are effectively filtered out.

We examine spatially connected regions formed by the retained responses and select the
region with the largest aggregated signal energy as the touch candidate. Based on this
region, the touch location is estimated using an intensity-weighted centroid:
\begin{equation}
(x_t, y_t) =
\frac{\sum_{(i,j)\in\mathcal{C}_t} (j,\, i)\, X_t(i,j)}
     {\sum_{(i,j)\in\mathcal{C}_t} X_t(i,j)},
\end{equation}
where $\mathcal{C}_t$ denotes the selected high-response region, and $x_t$ and $y_t$
correspond to the column and row coordinates, respectively.
The frame-wise touch positions $(x_t, y_t)$ are
subsequently refined using a constant-velocity Kalman filter with state
$\mathbf{s}_t = [x_t,\, y_t,\, v_{x,t},\, v_{y,t}]^\top$, yielding stable touch location
estimates for subsequent processing.

To better capture the spatial correlations of touch behaviors across frames, we flatten all capacitive frames corresponding to a single touch event. Specifically, the two-dimensional capacitive response of each frame is unfolded into a one-dimensional vector following a fixed ordering and concatenated over time, thereby representing a touch event as a continuous temporal feature sequence.
Based on this representation, we further apply temporal smoothing to the flattened sequence to mitigate frame-to-frame noise and local fluctuations, resulting in a more stable and continuous representation of touch behavior for subsequent analysis.

\subsubsection{Motion Estimation}
% TODO：修改，大致内容为：IMU的采样率和电容屏采样率差距过大，往往有20倍（引用论文），这会导致跨模态时间细粒度XXX的问题。因此，我们提出了XXX算法，大致就是先根据电容屏的跟踪算法定位粗区间，然后提出一个轻量化的IMU定位算法找到细区间，然后IMU的细区间和电容屏的细区间做时间域的对齐。然后这里再放一张对比图，关于电容数据的能量变化和IMU数据折线图，对比下对齐前和对齐后数据的一致性。修改完这块后记得修改系统框架图。
%TODO：在这小节最后加上不同用户频谱图的对比，以证明我们预处理的有效性，单栏
In this section, we preprocess raw IMU data to derive informative representations of touch behaviors. We first apply wavelet-based denoising to smooth the signals, followed by interpolation. The proposed refinement then extracts fine-grained IMU intervals corresponding to touch events. Quaternion-based rotation estimation is subsequently performed to capture rotational dynamics. Finally, the signals are transformed into the time-frequency domain via Short-Time Fourier Transform (STFT) to obtain spectral representations.

The IMU and the capacitive touchscreen operate at significantly different sampling rates, often differing by more than an order of magnitude. This discrepancy breaks the intrinsic coupling between touch events and their induced motion responses, leading to incorrect temporal correspondence.
To address this issue, we propose a lightweight coarse-to-fine cross-modal temporal refinement strategy.
We first use capacitive touch tracking to obtain a coarse temporal window $[t_i^s, t_i^e]$ for each touch event. Within this window, we refine the corresponding IMU interval using the acceleration magnitude $m(t)=|\mathbf{a}(t)|$. Specifically, we locate the dominant motion response by selecting the peak--valley pair with the maximum amplitude difference:
\begin{equation}
(t_i^{p}, t_i^{v}) = \arg\max_{t_p, t_v \in [t_i^s, t_i^e]} \left| m(t_p) - m(t_v) \right|.
\end{equation}

Since the extremum pair reflects the strongest motion response rather than the true temporal boundaries, we refine the interval locally around $(t_i^{p}, t_i^{v})$ by thresholding the first-order derivative of $m(t)$: the start point is defined as the first time the derivative exceeds a threshold when backtracking from $t_i^{p}$, and the end point as the first time it falls below the threshold when forward tracking from $t_i^{v}$.
The resulting fine-grained IMU interval is then aligned with the corresponding capacitive segment in the time domain.

Based on the aligned IMU segments, we further extract motion representations for subsequent modeling. In particular, to capture rotational dynamics, we derive orientation quaternions from the denoised accelerometer, gyroscope, and magnetometer measurements $(\boldsymbol{a}, \boldsymbol{g}, \boldsymbol{m})$, where $\boldsymbol{a}, \boldsymbol{g}$ and $\boldsymbol{m}$ denote the tri-axial acceleration, angular velocity, and magnetic-field vectors, respectively. 
The gyroscope’s angular velocity is first integrated through quaternion kinematics to obtain a preliminary orientation estimate. Subsequently, the accelerometer-derived gravity vector and the magnetometer-derived magnetic-field vector are incorporated to correct integration drift via a sensor-fusion update. This procedure yields a normalized orientation quaternion 
$
q_t = (q_0, q_1, q_2, q_3),
$
which forms the basis for deriving the rotation angles used in our feature-extraction and fusion pipeline. From this quaternion, we subsequently compute the corresponding Euler angles—roll ($\phi$), pitch ($\theta$), and yaw ($\psi$)—using the standard conversion formulas:
\begin{equation}
\phi = \arctan\left( \frac{2(q_2 q_3 + q_0 q_1)}{1 - 2(q_1^2 + q_2^2)} \right), 
\end{equation}

\begin{equation}
\theta = \arcsin\left( 2(q_1 q_3 - q_0 q_2) \right), 
\end{equation}

\begin{equation}
\psi = \arctan\left( \frac{2(q_1 q_2 + q_0 q_3)}{1 - 2(q_2^2 + q_3^2)} \right). 
\end{equation}
These angular features provide a stable and concise representation of the user's fine-grained touch-induced motion, enabling downstream feature extraction and multimodal fusion.

In addition to the orientation angles $(\phi, \theta, \psi)$, we incorporate the tri-axis acceleration components $(a_x, a_y, a_z)$ to more comprehensively capture subtle translational dynamics induced by touch interactions. 
By jointly modeling $(a_x, a_y, a_z, \phi, \theta, \psi)$ together with the extracted capacitive touchscreen touch-behavior data, we form a unified input sample that is fed into the \textit{Feature Extraction} module.

\subsection{Feature Extraction}
After constructing shallow representations of users’ touch behavior features, we further preprocess the data and train a feature extractor to extract deep representations for subsequent user authentication.

\begin{figure}[t]
    \centering
    \captionsetup{skip=0pt}
    \captionsetup[subfigure]{skip=0pt}
    \begin{subfigure}
    {\linewidth}
        \centering
        \includegraphics[width=\linewidth]{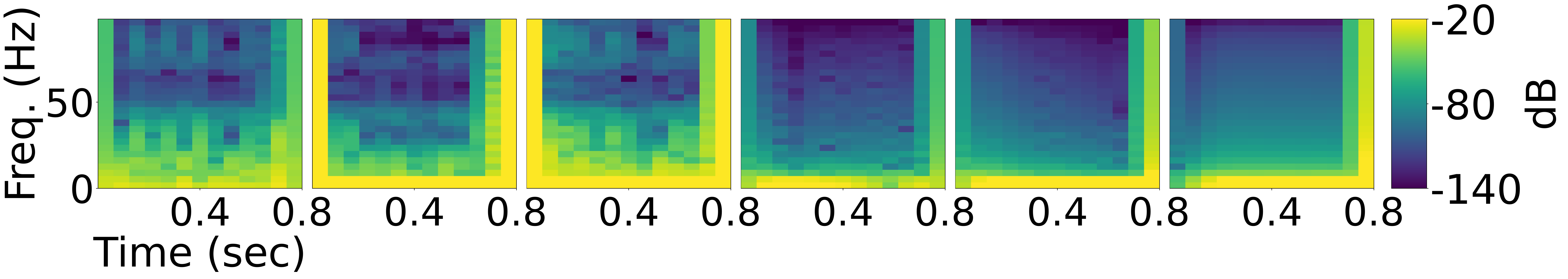}
    \end{subfigure}

    \begin{subfigure}{\linewidth}
        \centering
        \includegraphics[width=\linewidth]{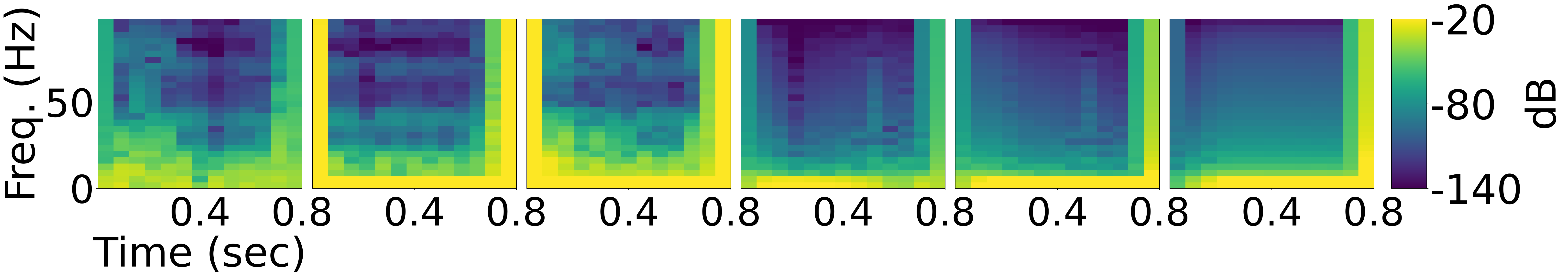}
        \caption{Two samples of User A.}
    \end{subfigure}

    \vspace{5pt}
    
    \begin{subfigure}{\linewidth}
        \centering
        \includegraphics[width=\linewidth]{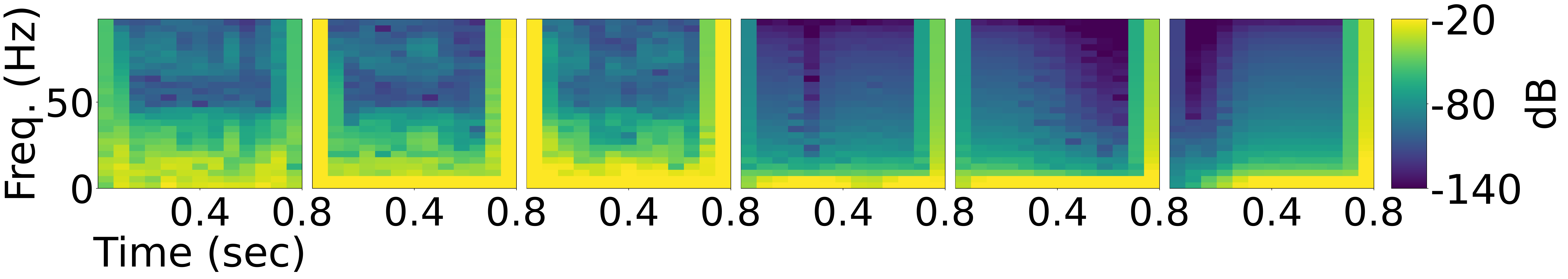}
    \end{subfigure}

    \begin{subfigure}{\linewidth}
        \centering
        \includegraphics[width=\linewidth]{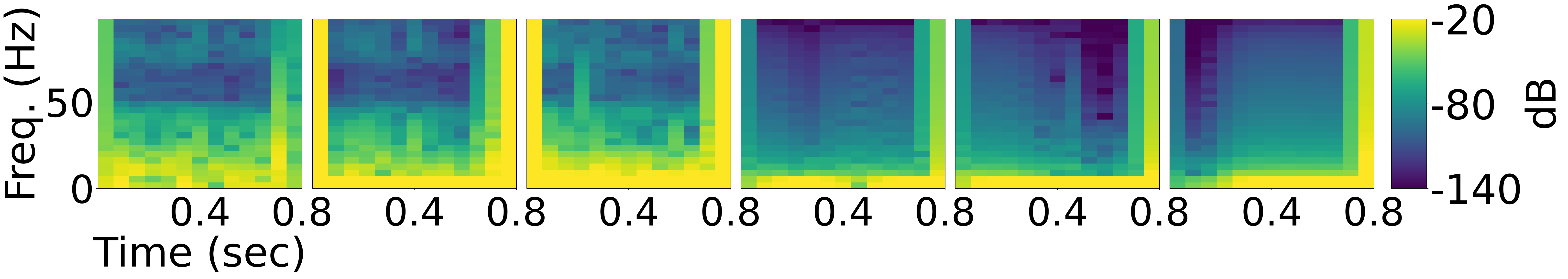}
        \caption{Two samples of User B.} 
    \end{subfigure}
    
    \vspace{8pt}
    
    \caption{Characterized touch behaviors of two users under STFT. Each row corresponds to one sample. From left to right, spectrograms of $a_x, a_y, a_z, \phi, \theta, \psi$.}
    \label{fig:spectrograms}

\end{figure}

We first perform time-frequency analysis on the IMU signals $(a_x, a_y, a_z, \phi, \theta, \psi)$ to obtain a fine-grained representation of user behavior from the IMU modality. 
Specifically, we apply the short-time Fourier transform (STFT) to each signal and convert them into two-dimensional power spectral density (PSD) matrices. 
These PSD matrices are concatenated to form an initial representation of the user’s IMU-side behavioral characteristics, which serves as the input to the subsequent multi-class classification model.
Fig.~\ref{fig:spectrograms} illustrates the characterized touch behaviors of two users under STFT, where two samples are presented for each user. From left to right, the spectrograms correspond to $a_x, a_y, a_z, \phi, \theta,$ and $\psi$, respectively.
Notably, samples from the same user exhibit strong consistency in their spectral patterns, while clear distinctions can be observed across different users. This demonstrates that the extracted representations effectively capture both intra-user consistency and inter-user discriminability.

For the capacitive touchscreen data, we first apply data augmentation to improve robustness against temporal variations and sensor noise. Specifically, time-axis warping is applied to the capacitive frame sequences with a warping factor of 0.1, corresponding to a random temporal stretching or compression of up to ±10\%, which simulates natural variations in touch speed and interaction rhythm.
In addition, amplitude-adaptive Gaussian noise is injected into individual capacitive frames. The noise standard deviation is proportional to the signal amplitude, with a base standard deviation of 0.5 and a minimum threshold of 0.1 to prevent insufficient perturbation in low-amplitude regions. The reference amplitude is estimated as the median of non-zero values within the central $3\times3$ region of the capacitive matrix, enabling the noise magnitude to adapt to different touch intensities. These augmentations introduce realistic perturbations while preserving the overall touch patterns.
The augmented capacitive touchscreen data are then fed into a separate multi-class classification model.
Notably, both the IMU-based model and the capacitive touchscreen-based model adopt the same network architecture, which is built upon a TinyViT~\cite{Wu2022TinyViT} backbone.
% TODO：在这里加上特征空间区分图

\begin{figure}[t]
  \centering
  \includegraphics[width=0.85\linewidth]{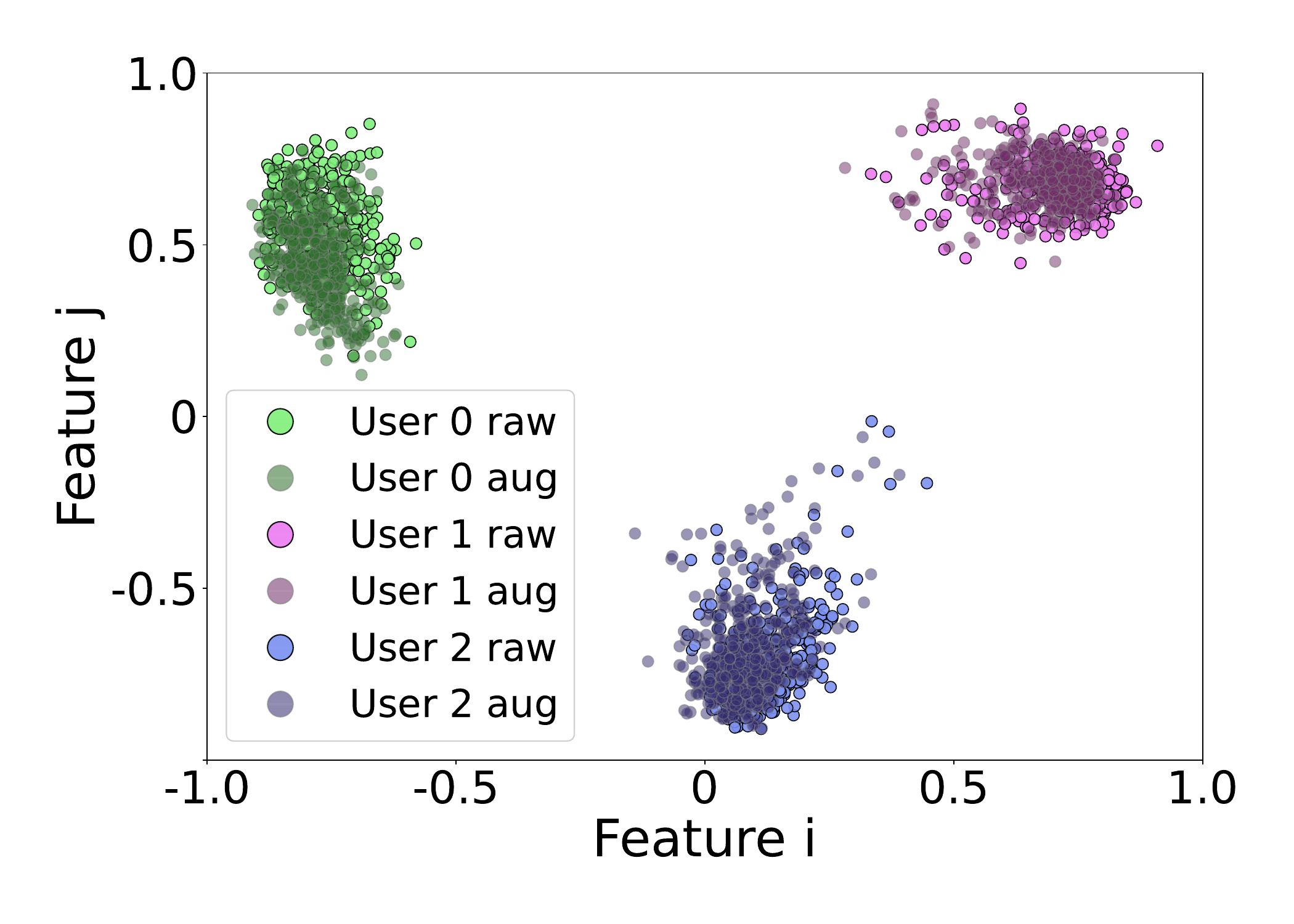} 
  \caption{Visualized feature space of raw and augmented (aug.) touch interaction data under PCA.} 
  \label{fig:pca} 
  \vspace{-2pt}
\end{figure}

Leveraging the transfer learning capability of deep networks, we remove the classification layers of the two trained multi-class models and repurpose their remaining backbones as feature extractors to obtain high-level, modality-specific representations. Based on these extracted features, we employ a separate network to further re-model and fuse information across modalities, which is trained independently for feature fusion.
The feature fusion network is implemented as a lightweight multilayer perceptron (MLP) consisting of two fully connected layers, with a LeakyReLU activation and a Dropout rate of 0.3. The network outputs a 320-dimensional feature vector, which serves as the final fused representation.
Fig.~\ref{fig:pca} presents a two-dimensional visualization of the original and augmented touch interaction features for three representative users. In both cases, samples from the same user form compact clusters with clear separation across users, indicating that the proposed framework extracts stable and robust behavioral representations.

\subsection{User Authentication}
In practical authentication scenarios, the training set typically contains samples from legitimate users only. Accordingly, the problem can be defined as a one-class classification task, in which an individual classifier is trained for each user. Specifically, each one-class classifier is trained using the 320-dimensional feature vectors extracted by the feature extractor.

We consider three widely used methods for profiling legitimate user behavior: i) One-Class Support Vector Machine (OC-SVM), which learns a decision boundary enclosing normal samples; ii) Local Outlier Factor (LOF), which identifies anomalies based on local density deviations; iii) Isolation Forest (IF), which 
isolates anomalies by randomly partitioning data.
We optimize the hyperparameters of the one-class classifiers using grid search and compare the performance of different methods in the Evaluation Section. Notably, the feature extraction and fusion models only require pretraining once and can be directly transferred to unseen subjects for inference.

\section{Data Collection}

\begin{figure}[t]
  \centering
  \includegraphics[width=0.6\linewidth]{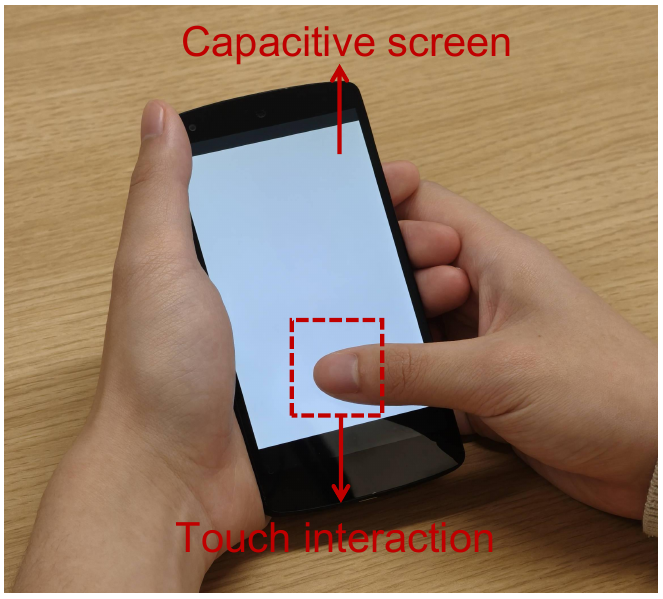} 
  %\vspace{-4pt}
  \caption{Illustration of the data collection process.} 
  \label{fig:data_collect} 
  \vspace{-2pt}
\end{figure}

To collect experimental data, we developed a prototype system on Android 9.0 (API level 28). Specifically, the system accesses capacitive touchscreen data through a wrapper built upon LibFTSP~\cite{huyle_capacitive}, with the touchscreen sampling rate set to 20 fps at a spatial resolution of 27 × 15. Meanwhile, the IMU signals are sampled at 200 Hz.
After obtaining IRB approval, we recruited 38 volunteers (14 females and 24 males), with ages ranging from 21 to 60 years. During formal data collection, participants held the device naturally with both hands and pressed a designated region on the touchscreen using a finger. 
An illustration of the data collection procedure is shown in Fig.~\ref{fig:data_collect}.
Each data collection session lasted 0.8 s, resulting in approximately 41,400 data points in total. Based on the collected data, we constructed 10 datasets for subsequent evaluation, as summarized in Table~\ref{Tab:Dataset}.

\begin{table*}[t]
\centering
\caption{Summary of Datasets Used in Evaluation.}
\label{Tab:Dataset}
\renewcommand{\arraystretch}{1.2}
\setlength{\tabcolsep}{4pt}

\begin{tabular}{lccll}
\toprule
\textbf{Dataset} & \textbf{\#Subjects} & \textbf{\# Samples} & \textbf{Setting / Environment} & \textbf{Experiment / Purpose} \\
\midrule
Dataset-1: Pre-training
& 38 & 15,200 & Seated; thumb press; 0.8\,s per press & Feature extractor training \\

Dataset-2: Auxiliary Authentication 
& 17 & 3,400 & Seated; PIN input (4578); fingerprint unlock & Auxiliary authentication analysis \\

Dataset-3: Mimicry Attack
& 8 pairs & 800 & Seated; attacker imitation & Mimicry attack evaluation \\

Dataset-4: Artificial Attack
& 8 & 800 & Seated; fingerprint replica; thumb press & Artificial replication attack evaluation\\

Dataset-5: Puppet Attack
& 8 pairs & 800 & Seated; attacker-controlled victim finger & Puppet attack evaluation\\

Dataset-6: Overtime
& 8 & 4,000 & Seated; Weeks 1-5 after Dataset-1 & Long-term stability evaluation \\

Dataset-7: Cross-finger
& 8 & 9,600 & Seated; thumb / index / middle finger press & Cross-finger generalization \\

Dataset-8: Motion-state
& 8 & 1,200 & Standing / lying / walking; thumb press & Motion-state robustness \\

Dataset-9: Moisture
& 8 & 3,200 & Seated; dry vs. wet finger & Moisture robustness \\

Dataset-10: Screen Protector
& 8 & 2,400 & Seated; hydrogel / PET / tempered glass film & Screen condition robustness \\
\bottomrule
\end{tabular}
\end{table*}

(1) \textit{Dataset-1.} 
The dataset is constructed from 38 subjects, each performing 400 valid thumb presses while sitting naturally, holding the device with both hands, and pressing a designated touchscreen region with their thumb—mimicking typical under-display fingerprint unlocking. This yields $38 \times 400 = 15{,}200$ samples in \textit{Dataset-1}. The data are split into three non-overlapping subsets: data from 15 subjects for feature extractor training, 23 subjects for user registration and authentication evaluation, and negative samples constructed from all subjects with strict exclusion of overlap with corresponding positive samples in each experiment.

(2) \textit{Dataset-2.} 
To evaluate our method as an auxiliary mechanism for both PIN-based and fingerprint-based authentication, \textit{Dataset-2} is constructed from 17 subjects under the default experimental setting. Subjects sit naturally, hold the device with both hands, and unlock the screen using a fixed PIN code (“4578”) to eliminate content-related variations. Each subject performs 100 valid PIN-based unlocking interactions. The same 17 subjects also perform fingerprint-based authentication under identical conditions, allowing evaluation of our method as a unified auxiliary signal across different primary authentication mechanisms. In total, \textit{Dataset-2} contains $17 \times 200 = 3{,}400$ auxiliary authentication samples.

(3) \textit{Dataset-3.}
We construct a mimicry attack scenario in which attackers attempt to imitate legitimate users with their own fingers, by visually observing the victim’s authentication process and mimicking the corresponding finger placement and pressing behaviors. %Each attacker is paired with one target subject and is allowed to closely observe the victim’s pressing posture, touch location, and force patterns before reproducing the authentication behavior.
A total of 8 attackers are randomly paired with 8 subjects selected from Dataset-1. Each attacker performs 100 mimicry attempts per subject under each imitation setting, yielding $8  \times 100 = 800$ samples.

\begin{figure}[t]
  \centering
  \includegraphics[width=1.0\linewidth]{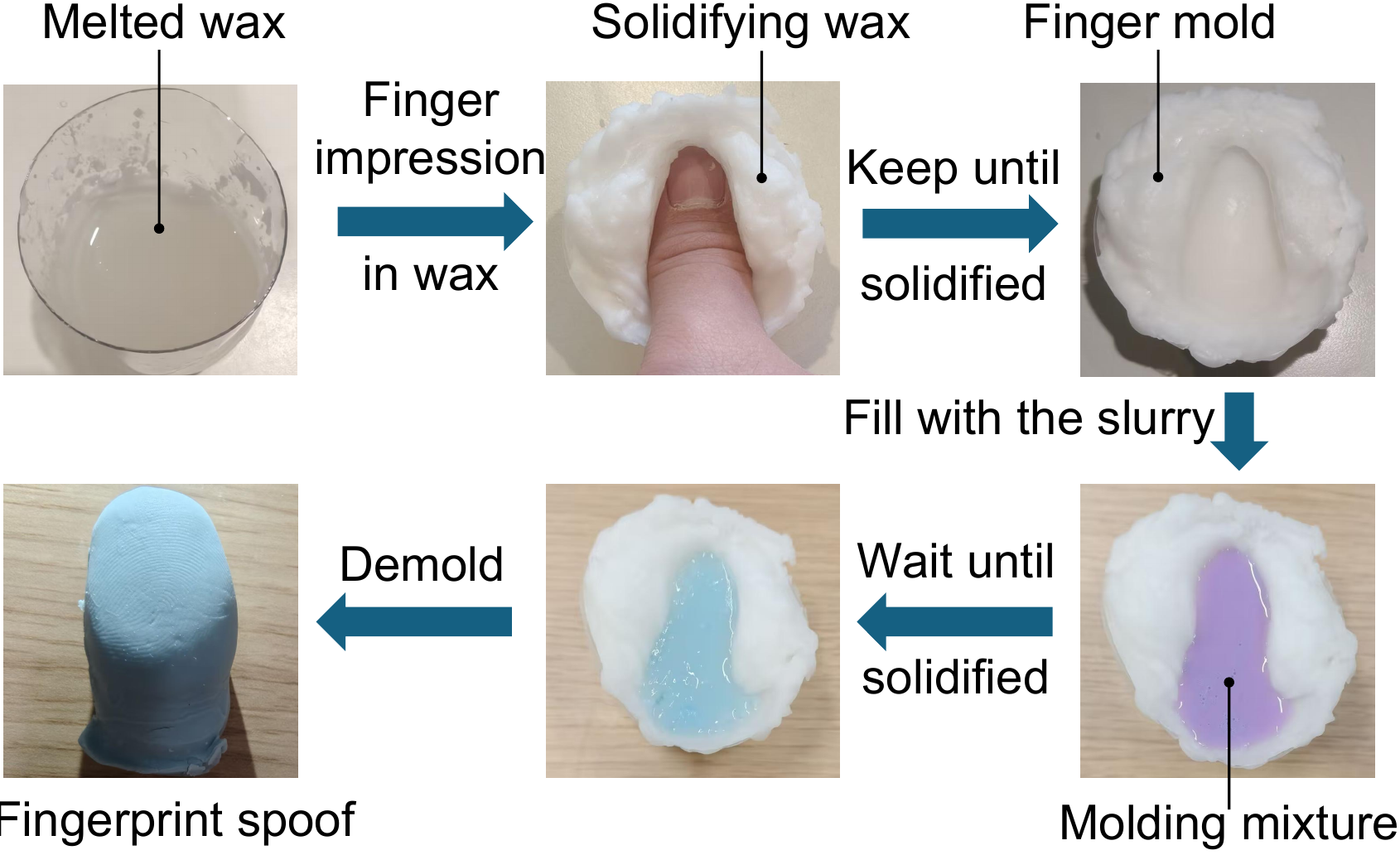} 
  %\vspace{-4pt}
  \caption{Fabrication procedure of fingerprint spoofs.} 
  \label{fig:fingerspoof} 
  \vspace{-5pt}
\end{figure}

(4) \textit{Dataset-4.}
To evaluate artificial replication attacks, \textit{Dataset-4} is constructed using fingerprint spoofs fabricated for 8 subjects. As illustrated in Fig.~\ref{fig:fingerspoof}, the fingerprint spoofs are created using a mixture of sodium alginate and plaster powder, which solidifies into soft and conductive fake fingers that can be sensed by capacitive touchscreens.
To ensure high geometric similarity to real fingers, the fabricated fingerprint spoofs are required to successfully pass verification by a professional fingerprint sensor (Live~20R) from ZKTeco. Each subject performs 100 thumb-press interactions, resulting in a total of $8 \times 100 = 800$ samples in \textit{Dataset-4}.

(5) \textit{Dataset-5.}
We construct a puppet attack scenario in which the attacker holds the device with one hand while physically manipulating the victim’s finger with the other to perform authentication. During the attack, the victim does not actively apply force, thereby simulating a non-consensual or incapacitated condition.
The attacker is allowed to closely observe the victim’s pressing behavior in advance and may apply arbitrary force when controlling the victim’s finger. For the 8 subjects randomly selected from \textit{Dataset-1}, the attacker performs 100 puppet attack attempts per subject, resulting in a total of $8 \times 100 = 800$ samples.

(6) \textit{Dataset-6.}
To evaluate authentication performance over time, we recollect press data from 8 subjects at multiple time points, specifically at Weeks 1, 2, 3, 4, and 5 after the initial data collection. At each time point, each subject performs 100 seated thumb-press interactions following the same default settings.
In total, \textit{Dataset-6} contains $8 \times 100 \times 5 = 4{,}000$ samples.

(7) \textit{Dataset-7.}
To evaluate the impact of different authentication fingers on system performance, we collect thumb-, index-, and middle-finger press data from 8 subjects. The data collection protocol follows the same default setting as Dataset-1, where subjects remain seated and hold the device naturally. For each subject, 400 valid presses are recorded for each finger.
As a result, \textit{Dataset-7} contains a total of $8 \times 3 \times 400 = 9{,}600$ samples.

(8) \textit{Dataset-8.}
To evaluate the impact of different authentication postures on system performance, we collect press data from 8 subjects under multiple motion states. Specifically, each subject performs thumb-press interactions while standing, lying down, and walking, following their habitual under-display fingerprint unlocking behavior. For each posture, 50 valid presses are recorded.
As a result, \textit{Dataset-8} contains a total of $8 \times 50 \times 3 = 1{,}200$ samples.

(9) \textit{Dataset-9.}
To evaluate the impact of finger moisture on authentication performance, we collect both dry- and wet-finger press data from 8 subjects. For each subject, 300 presses are performed with a dry finger and 100 presses with a wet finger, resulting in a controlled imbalance between the two conditions.
In total, \textit{Dataset-9} includes $8 \times (300+100) = 3{,}200$ samples.

(10) \textit{Dataset-10.}
To assess the impact of screen protectors on authentication performance, we evaluate three commercially available types: a tempered glass protector (Glass), a hydrogel-based TPU protector (Hydrogel), and a standard PET film protector (PET).
A total of 8 subjects participate in this experiment, and each subject performs 100 seated thumb-press interactions under each screen protector condition, following the default protocol.
As a result, \textit{Dataset-10} contains $8 \times 3 \times 100 = 2{,}400$ samples.
\begin{figure}[t]
  \centering
  \setlength{\tabcolsep}{2pt}
  \begin{subfigure}[t]{0.32\linewidth}
    \centering
    \includegraphics[width=\linewidth]{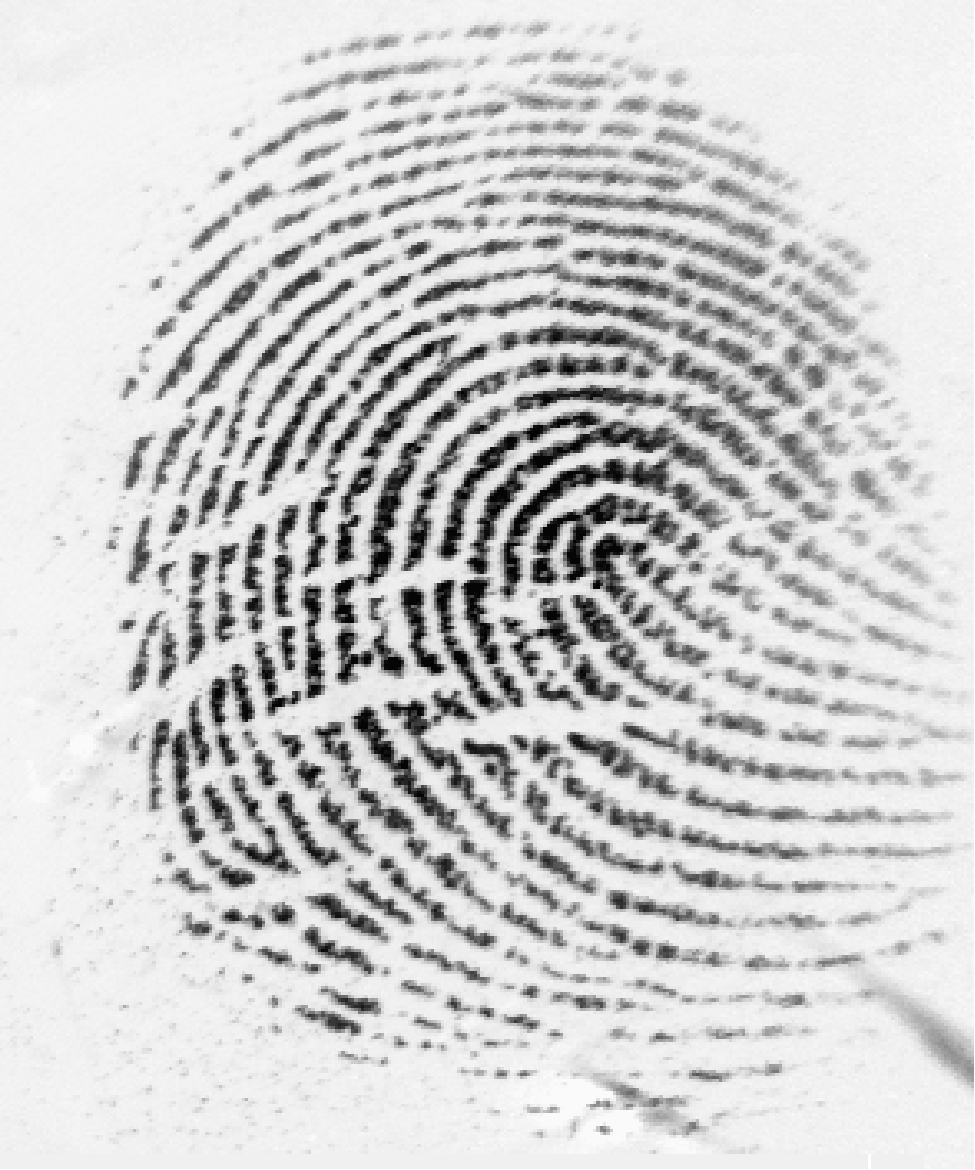}
    \caption{Genuine image}
  \end{subfigure}\hfill
  \begin{subfigure}[t]{0.32\linewidth}
    \centering
    \includegraphics[width=\linewidth]{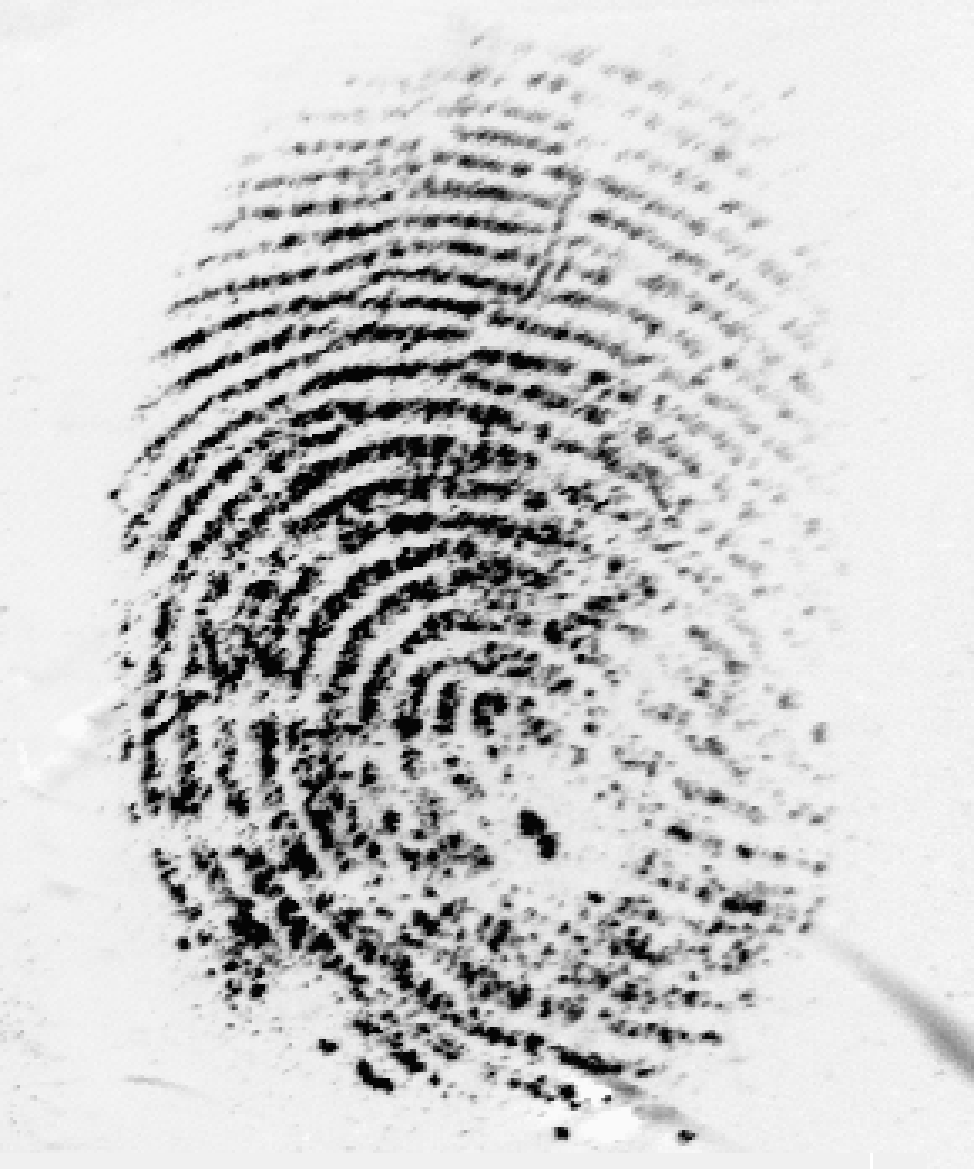}
    \caption{Spoofed image}
  \end{subfigure}\hfill
  \begin{subfigure}[t]{0.32\linewidth}
    \centering
    \includegraphics[width=\linewidth]{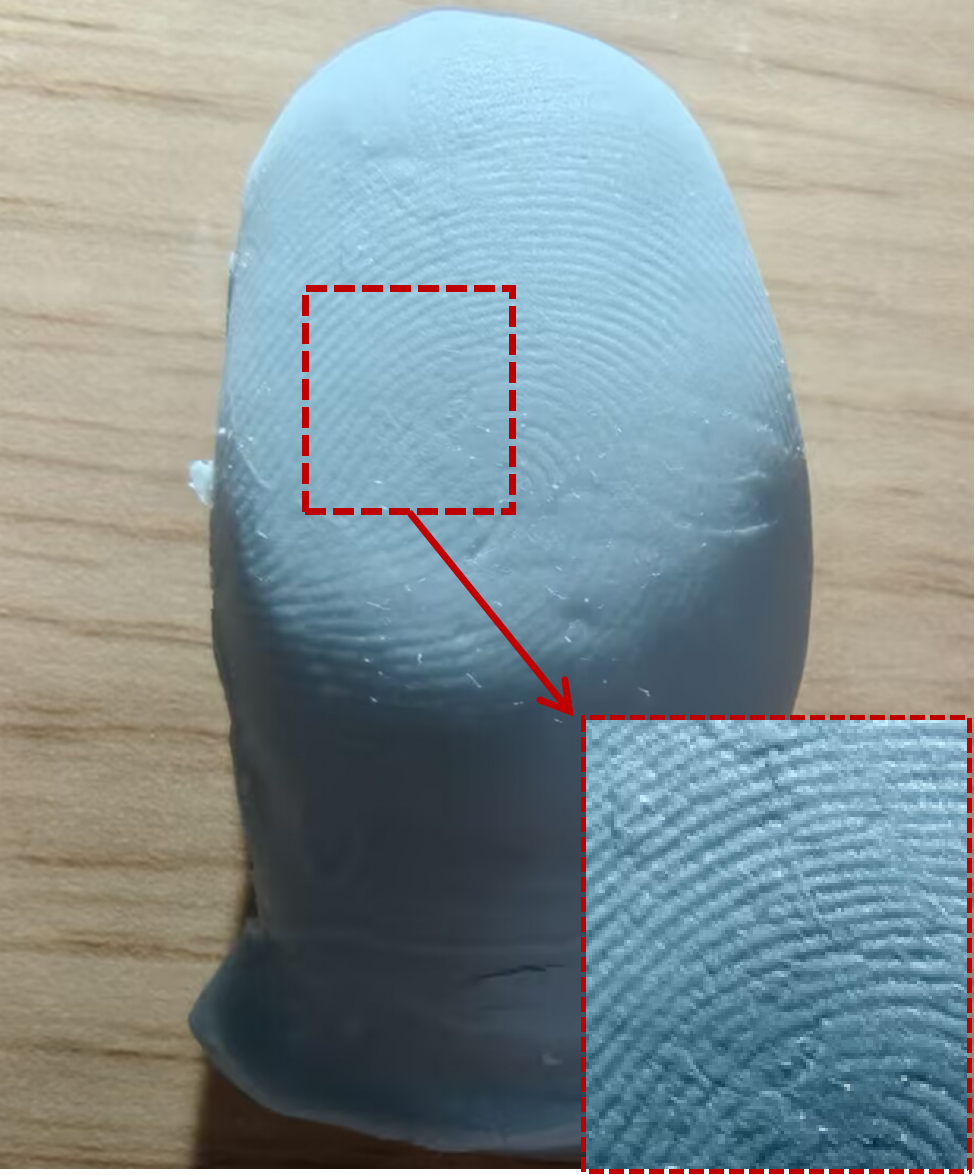}
    \caption{Spoof artifact}
  \end{subfigure}
  \caption{Example fingerprint images acquired by the fingerprint sensor, including a genuine fingerprint, a spoofed fingerprint impression, and the corresponding fabricated fingerprint spoof.}
  \label{fig:spoof_compare}
\end{figure}

\section{Evaluation}

\subsection{Experimental Settings}

\textbf{Default Setting.}
We use an LG Nexus 5 smartphone as the default data collection device. Participants are seated on a chair and instructed to press a designated region on the touchscreen using their finger.
In all experiments, the training and testing sets are constructed with a 1:1 sample ratio. Moreover, the training and testing data are strictly separated by participants, such that the model does not have prior access to any user data from the testing set.
For artificial replication attacks, we use a mixture of sodium alginate and plaster powder to fabricate fake fingers whose conductivity is similar to that of real human fingers. The fabricated fake fingers are required to successfully pass verification by the Live 20R fingerprint sensor from ZKTeco Live20R, ensuring high geometric similarity to real fingers.
Fig.~\ref{fig:fingerspoof} presents the fabrication procedure of the fake fingers, while Fig.~\ref{fig:spoof_compare} shows example fingerprint images acquired by the fingerprint sensor.

\textbf{Evaluation Metrics.}
We evaluate the authentication performance using the following metrics. False Accept Rate (FAR) measures the proportion of illegitimate authentication attempts that are incorrectly accepted and is defined as $\mathrm{FAR}=\mathrm{FA}/(\mathrm{FA}+\mathrm{TR})$, where FA and TR denote false accepts and true rejects, respectively.
False Reject Rate (FRR) measures the proportion of legitimate authentication attempts that are incorrectly rejected and is defined as $\mathrm{FRR}=\mathrm{FR}/(\mathrm{FR}+\mathrm{TA})$, where FR and TA denote false rejects and true accepts.
Accuracy represents the overall proportion of correctly classified samples and is defined as $\mathrm{Accuracy}=(\mathrm{TA}+\mathrm{TR})/(\mathrm{TA}+\mathrm{TR}+\mathrm{FA}+\mathrm{FR})$.
Balanced Accuracy (BAC) is used to account for class imbalance and is defined as the arithmetic mean of the true accept rate and the true reject rate, i.e., $\mathrm{BAC}=\frac{1}{2}\big(\mathrm{TA}/(\mathrm{TA}+\mathrm{FR})+\mathrm{TR}/(\mathrm{TR}+\mathrm{FA})\big)$.
Equal Error Rate (EER) refers to the operating point at which FAR equals FRR.

\begin{figure*}[t!]
    \begin{minipage}[t]{0.32\linewidth}
        \centering
        \includegraphics[width=\textwidth]{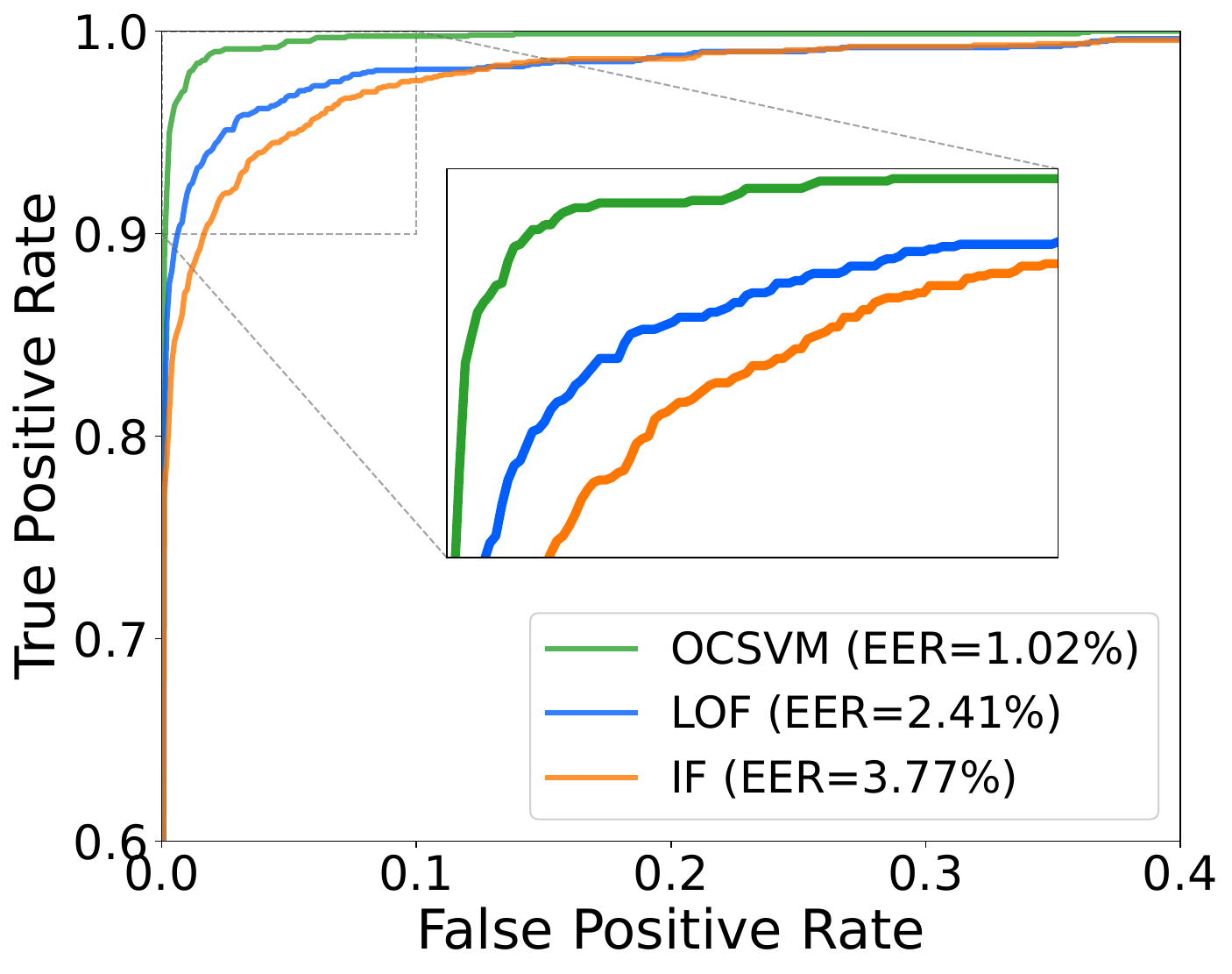} 
        \vspace{-2pt}
        \caption{ROC curves of the IMU-based method.} 
        \label{fig:roc_imu} 
    \end{minipage}
    \hfill   
    \begin{minipage}[t]{0.32\linewidth}
        \centering
        \includegraphics[width=\textwidth]{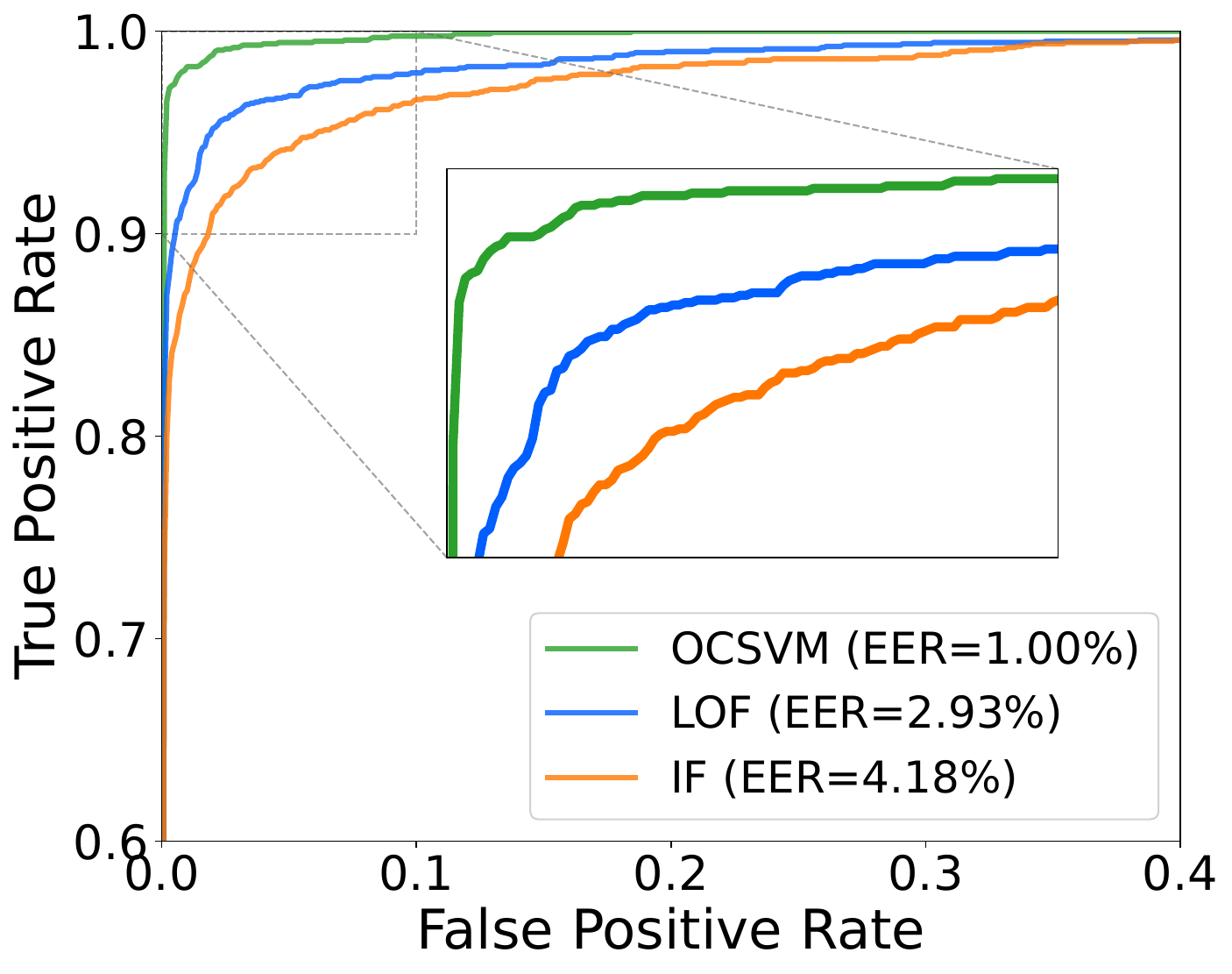}
        \vspace{-2pt}
        \caption{ROC curves of the capacitive-based method.}
        \label{fig:roc_cap}
    \end{minipage}
    \hfill
    \begin{minipage}[t]{0.32\linewidth}
        \centering
        \includegraphics[width=\textwidth]{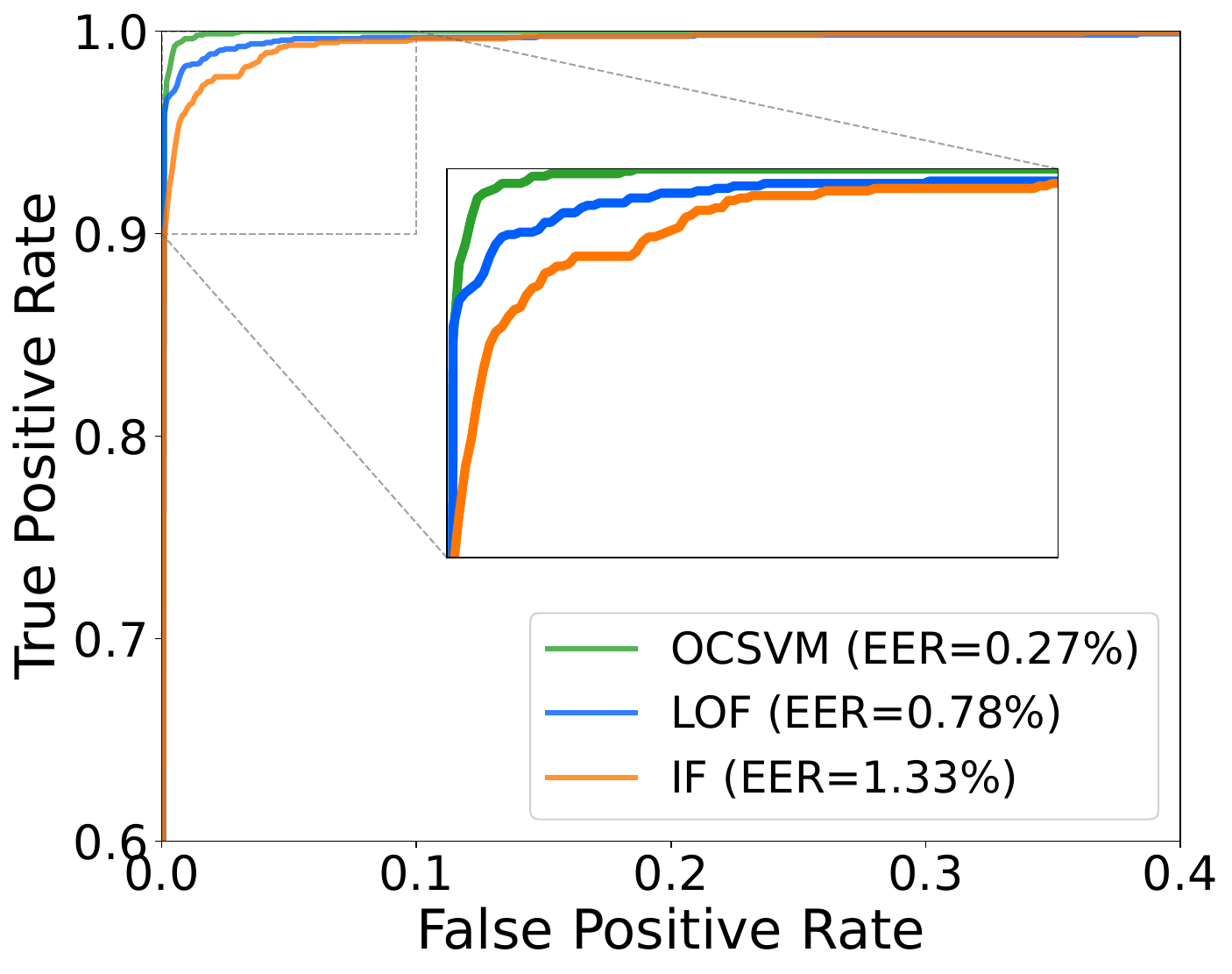}
        \vspace{-2pt}
        \caption{ROC curves of BioMoTouch.}
        \label{fig:roc_fusion}
    \end{minipage}
\end{figure*}

\begin{figure*}[t!]
    \begin{minipage}[t]{0.32\linewidth}
        \centering
        \includegraphics[width=\textwidth]{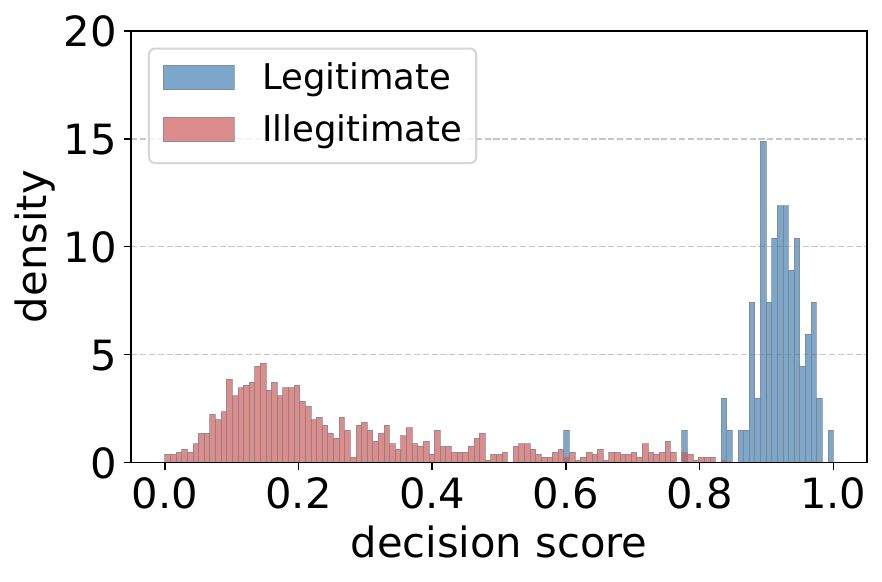} 
        \vspace{-2pt}
        \caption{Decision score distributions of the IMU-based method.} 
        \label{fig:decision_imu} 
    \end{minipage}
    \hfill   
    \begin{minipage}[t]{0.32\linewidth}
        \centering
        \includegraphics[width=\textwidth]{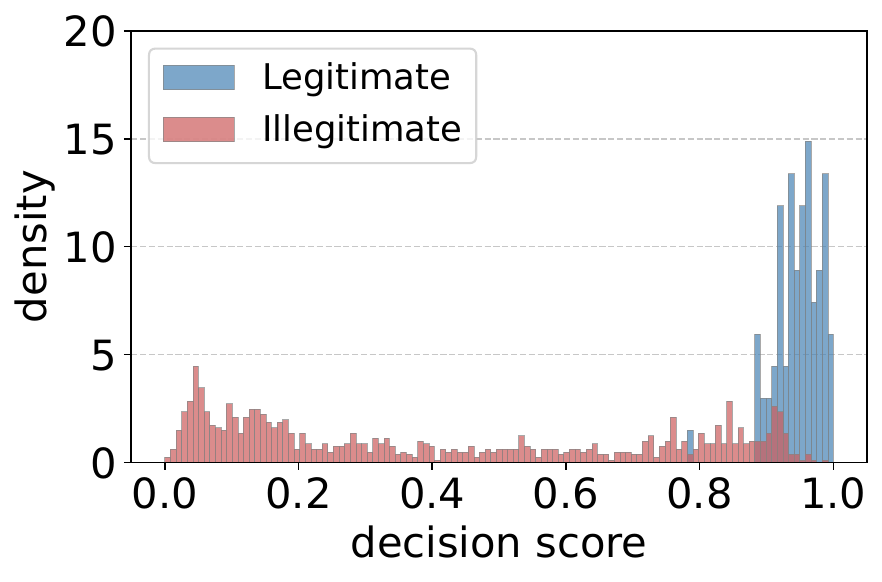}
        \vspace{-2pt}
        \caption{Decision score distributions of the capacitive-based method.}
        \label{fig:decision_cap}
    \end{minipage}
    \hfill
    \begin{minipage}[t]{0.32\linewidth}
        \centering
        \includegraphics[width=\textwidth]{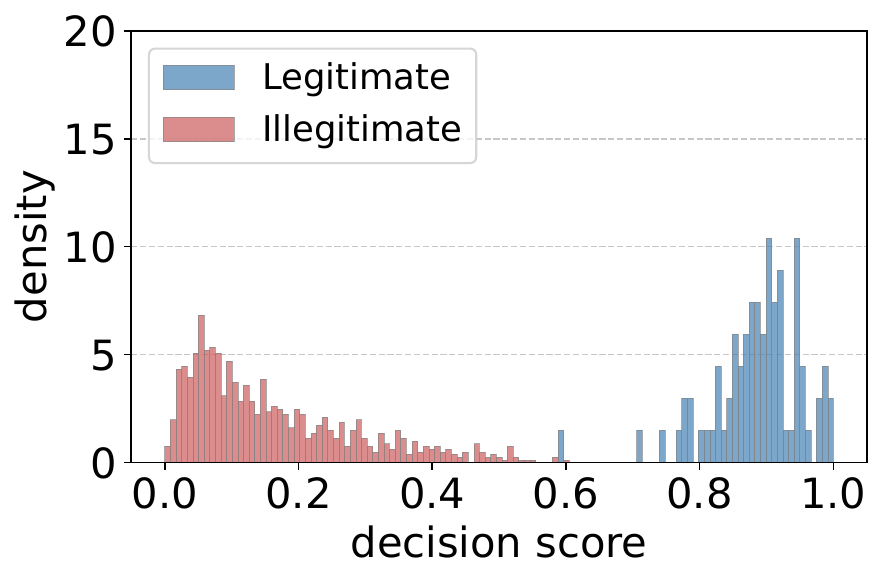}
        \vspace{-2pt}
        \caption{Decision score distributions of BioMoTouch.}
        \label{fig:decision_fusion}
    \end{minipage}
\end{figure*}

\subsection{Overall Performance}
To evaluate the overall authentication performance of the proposed system, we conduct a comprehensive set of experiments covering authentication effectiveness, computational efficiency, modality ablation, feature-classifier configurations, baseline comparisons, auxiliary authentication capability, and robustness against adversarial attacks. Together, these evaluations provide an overall assessment of the system under standard authentication settings.

\subsubsection{Authentication Performance}
To evaluate the effectiveness of BioMoTouch, we train and evaluate authentication models using \textit{Dataset-1}. As shown in Fig.~\ref{fig:roc_fusion}, BioMoTouch demonstrates consistently strong discriminative performance across all classifiers. In particular, the ROC curve of the OC-SVM variant lies closest to the upper-left corner, indicating a favorable trade-off between false positive rate and true positive rate. Compared to LOF and IF, OC-SVM achieves the lowest EER (0.27\%), highlighting its superior capability in modeling user-specific touch patterns. The clear separation among the curves further confirms the effectiveness of the proposed multimodal representation for authentication.

\subsubsection{Computational Efficiency Analysis}
We analyze the computational efficiency of BioMoTouch on a workstation equipped with an Intel(R) Core(TM) i7-12900K processor, without using GPU acceleration, in order to approximate the computational constraints of portable intelligent terminals such as smartphones. Under this CPU-only setting, a single authentication operation incurs an average latency of approximately 44.6 ms, including 13.502 ms for preprocessing and 29.938 ms + 184.99 $\mu s$ for feature extraction and one-class classification.
These results indicate that BioMoTouch introduces only limited computational overhead and can support real-time authentication under resource-constrained settings, demonstrating its practicality for deployment on commodity mobile devices.

\subsubsection{Ablation Study of Sensing Modalities}
To empirically validate that capacitive sensing alone can support identity discrimination and provide evidence of physiological information embedded in the touch interaction, and examine the contribution of each sensing modality in BioMoTouch, we conduct an ablation study using \textit{Dataset-1} by decomposing the system into an IMU-based variant and a capacitive-screen-based variant. The corresponding ROC curves are shown in Figs.~\ref{fig:roc_imu}--\ref{fig:roc_fusion}.
As illustrated in Fig.~\ref{fig:roc_imu}, the IMU-based method exhibits limited discriminative capability, with EERs of 1.02\% (OC-SVM), 2.41\% (LOF), and 3.77\% (IF), indicating relatively high false acceptance and rejection rates when relying solely on inertial signals. Fig.~\ref{fig:roc_cap} shows that the capacitive-screen-based method achieves improved performance compared to the IMU-based variant, reducing the EERs to 1.00\%, 2.93\%, and 4.18\%, respectively.
Importantly, although capacitive sensing alone does not achieve optimal performance, it already exhibits measurable discriminative capability. This observation supports our empirical finding that natural touch interaction encodes user-specific physiological characteristics that can be directly captured by commodity capacitive screens.
In contrast, BioMoTouch in Fig.~\ref{fig:roc_fusion} exhibits the most favorable ROC characteristics across all classifiers, achieving lower EERs of 0.27\% (OC-SVM), 0.78\% (LOF), and 1.33\% (IF).

To further analyze the limitations of single-modality approaches, we examine the decision score distributions of legitimate and illegitimate samples produced by the IMU-based and capacitive-screen-based variants, as shown in Figs.~\ref{fig:decision_imu}--\ref{fig:decision_fusion}.
For both variants, the two classes exhibit noticeable overlap in decision scores, indicating limited separability under single-modality sensing. While capacitive sensing captures physiological characteristics embedded in touch interaction, it does not fully characterize the associated behavioral dynamics of each touch event. Similarly, IMU signals reflect motion-related behavioral patterns but lack detailed physiological information. As a result, modeling either dimension in isolation leads to insufficient separation between legitimate and illegitimate samples.
In contrast, BioMoTouch markedly reduces this overlap, yielding more separable decision-score distributions. This result validates that explicitly modeling the intrinsic coupling between physiological structure and behavioral dynamics improves discriminative robustness under realistic conditions.

\begin{table*}[t]
\centering
%\scriptsize
\caption{Comparison with Representative Commercial and Research Authentication Systems.}
\label{tab:commercial_compare}
\renewcommand{\arraystretch}{1.2}
\setlength{\tabcolsep}{6pt}
\begin{tabular}{l l c c c c c r}
\toprule
\textbf{Method} &
\textbf{Description of features} &
\textbf{\makecell{No extra \\hardware}} &
\textbf{\makecell{Ambient \\robustness}$^1$} &
\textbf{\makecell{Replica \\resistance}} &
\textbf{\makecell{Puppet attack \\resistance}} &
\textbf{BAC $\uparrow$$^2$} \\
\hline
\multicolumn{8}{c}{Commercial products} \\
\hline

Samsung~\cite{samsung_ultrasonic_fp}  &
Ultrasonic fingerprint features &
\ding{55} &
\ding{51} &
\ding{51} &
\ding{55} &
N/A \\

Qualcomm~\cite{qualcomm}  &
Ultrasonic fingerprint features &
\ding{55} &
\ding{51} &
\ding{51} &
\ding{55} &
N/A \\

BehavioSec~\cite{behaviosec} &
Touch dynamics (trajectory, pressure, timing) &
\ding{51} &
\ding{51} &
\ding{51} &
\ding{55} &
N/A \\

BioCatch~\cite{biocatch} &
Behavioral biometrics &
\ding{51} &
\ding{51} &
\ding{51} &
\ding{55} &
N/A \\

ZKTeco Live20R~\cite{zkteco_live20r} &
Fingerprint image features &
\ding{55} &
\ding{51} &
\ding{55} &
\ding{55} &
N/A \\

Apple Touch ID~\cite{apple_touch_id}&
Capacitive fingerprint image features &
\ding{55} &
\ding{51} &
\ding{51} &
\ding{55} &
N/A \\

\hline
\multicolumn{8}{c}{Research paper} \\
\hline

TouchPrint~\cite{chen2020listentofingers} &
Hand posture shape traits & %Hand geometry acoustic features &
\ding{51} &
\ding{55} &
\ding{55} &
\ding{55} &
91.7\% \\

TouchPass~\cite{xu2020touchpass} &
Cepstrum-based vibration features &
\ding{51} &
\ding{51} &
\ding{55} &
\ding{55} &
93.5\% \\

FingerSlid~\cite{10251599} &
Sliding touch dynamics &
\ding{51} &
\ding{51} &
\ding{51} &
\ding{55} &
95.4\% \\

Fingerbeat~\cite{10443592} &
Sliding touch dynamics &
\ding{51} &
\ding{51} &
\ding{51} &
\ding{55} &
95.1\% \\

PressPIN~\cite{9714878} &
Tap-induced acoustic features &
\ding{51} &
\ding{55} &
\ding{55} &
\ding{55} &
$\sim$ 83.0\% \\

MMAuth~\cite{9737094} &
Touch-based hand motion trajectories &
\ding{51} &
\ding{55} &
\ding{51} &
\ding{51} &
$\sim$ 85.0\% \\

FingerVib~\cite{Wu2025FingerVib} &
motion dynamics and audio-assisted vibration &
\ding{51} &
\ding{55} &
\ding{51} &
\ding{51} &
98.4\% \\

\textbf{BioMoTouch} &
\textbf{Touch behavior and hand motion dynamics} &
\ding{51} &
\ding{51} &
\ding{51} &
\ding{51} &
\textbf{99.7\%}
 \\
\bottomrule
\end{tabular}
\begin{minipage}{0.98\linewidth}
\footnotesize
$^1$ indicates that authentication performance remains stable across variations in user posture, physical environment, and over time. \\
$^2$ indicates the balanced authentication accuracy (BAC) for previously unseen users based on a single authentication attempt.

\end{minipage}
\end{table*}

\begin{table}[t]
\centering
\caption{Performance of Different Feature Extractors Combined with One-Class Classifiers.}
\label{tab:oneclass_combination}
\setlength{\tabcolsep}{2pt}
\renewcommand{\arraystretch}{1.5}
\begin{tabular}{lccc}
\hline
\textbf{\makecell{Feature \\ Extractor}} &
\textbf{\makecell{OC-SVM \\ (BAC $\uparrow$ / EER $\downarrow$)}} &
\textbf{\makecell{LOF \\ (BAC $\uparrow$ / EER $\downarrow$)}} &
\textbf{\makecell{IF \\ (BAC $\uparrow$ / EER $\downarrow$)}} \\
\hline
ResNet50~\cite{He_2016_CVPR}    &
98.87\% / 0.52\% & 98.83\% / 1.06\% & 97.81\% / 2.08\% \\
EfficientNet~\cite{Tan2019EfficientNet}   & 
98.60\% / 0.72\% & 99.02\% / 0.98\% & 96.55\% / 3.35\% \\
TinyViT~\cite{Wu2022TinyViT}   & 
\textbf{99.71\% / 0.27\%} & 99.03\% / 0.78\% & 98.45\% / 1.33\% \\
\hline
\end{tabular}
\end{table}

\subsubsection{Feature Extractor and One-Class Classifier Selection}
To evaluate the impact of different feature extractors under one-class classification settings, we evaluate several representative backbone models, including ResNet50~\cite{He_2016_CVPR}, EfficientNet~\cite{Tan2019EfficientNet}, and TinyViT~\cite{Wu2022TinyViT}, in combination with commonly used one-class classifiers, i.e., OC-SVM, LOF, and IF. 
Table~\ref{tab:oneclass_combination} reports the authentication performance of these feature extractor and classifier combinations. Overall, TinyViT consistently achieves superior performance across all evaluated one-class classifiers, and attains the highest BAC (99.71\%) and the lowest EER (0.27\%) when combined with OC-SVM, indicating its strong representation capability for modeling touch behavior. Based on these results, we adopt the TinyViT + OC-SVM combination as the default configuration in subsequent experiments.

\subsubsection{Baseline Comparisons}
Existing authentication systems can be broadly categorized into commercial products and research-oriented prototypes based on their sensing modalities and interaction designs. Table~\ref{tab:commercial_compare} summarizes a comparison between representative mature commercial systems and recent research-based methods.
Commercial authentication products, including ultrasonic and capacitive fingerprint sensors~\cite{samsung_ultrasonic_fp,qualcomm,zkteco_live20r,apple_touch_id}, rely on static fingerprint patterns feature, as well as touch dynamics apps~\cite{behaviosec,biocatch}, which rely on coarse behavioral signals. These systems typically require dedicated hardware and provide limited resistance to advanced attacks, particularly lacking robustness against puppet attacks.
Recent research efforts have explored touch- and motion-based authentication using commodity sensors, such as capacitive touchscreens, microphones, accelerometers, and IMUs~\cite{chen2020listentofingers,xu2020touchpass,10251599,10.1145/3133956.3133964,9714878,9737094,Wu2025FingerVib}. These methods achieve authentication accuracies ranging from approximately 83.0\% to 98.4\% under a single authentication attempt, but many of them rely on behavioral patterns or contact-induced responses, and exhibit limited robustness under posture or environmental variations, as well as under replica and puppet attack scenarios.
In contrast, BioMoTouch leverages only capacitive touchscreen and IMU sensors and operates transparently during natural user interactions. As shown in Table~\ref{tab:commercial_compare}, it achieves the highest authentication accuracy among research-based approaches (99.7\%) while simultaneously requiring no additional hardware, maintaining robustness under diverse conditions, and providing resistance to both replica and puppet attacks.

\subsubsection{Performance of Auxiliary Authentication}
To evaluate the effectiveness of BioMoTouch as an auxiliary authentication mechanism, we conduct experiments on \textit{Dataset-2}, which includes both PIN-based and fingerprint-based unlocking interactions collected under identical conditions. BioMoTouch is evaluated as an additional behavioral authentication signal alongside the primary authentication mechanism.
For fingerprint-based authentication, BioMoTouch achieves an EER of 0.27\% when used as an auxiliary signal. For PIN-based authentication, the EER is further reduced to 0.19\%. These EER values are computed solely from BioMoTouch’s decision scores, independent of the primary authentication mechanisms, indicating that touch interaction behaviors remain highly consistent across different primary authentication settings.

\begin{figure}[t]
  \centering
  \includegraphics[width=0.7\linewidth]{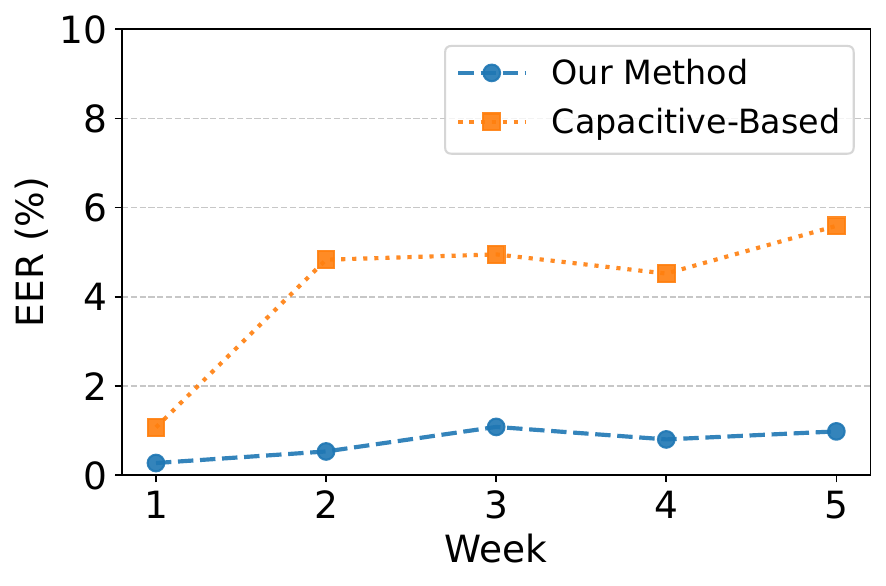} 
  \caption{Long-term EER comparison over five weeks.} 
  \vspace{17pt}
  \label{fig:overtime} 
\end{figure}

\subsubsection{Attack Resistance Evaluation}
To evaluate the robustness of BioMoTouch against adversarial behaviors, we conduct experiments on \textit{Dataset-3}, \textit{Dataset-4}, and \textit{Dataset-5} to compare our method with a representative commercial solution, ZKTeco Live20R, under three common attack scenarios: Mimicry Attack, Artificial Replication Attack, and Puppet Attack. As reported in Table~\ref{tab:far_attacks}, ZKTeco Live20R exhibits relatively high false acceptance rates under these attacks, with FAR reaching 15.73\% for artificial replication and 100\% under the puppet attack.
In contrast, the proposed method consistently achieves low FAR across all three attack scenarios, with values of 0.51\%, 0.90\% and 0.51\%, respectively. Overall, these results indicate that the proposed approach is more resilient to practical attack strategies than the commercial baseline.

\subsection{Reliability Analysis}
To evaluate the reliability of the proposed authentication system under realistic usage conditions, we conduct a systematic analysis of several practical factors that may affect authentication performance. 
In addition, we compare BioMoTouch with our capacitive-screen-based variant to assess the relative robustness of different sensing modalities under these practical variations.

\begin{table}[t]
\centering
\caption{False Acceptance Rate (FAR) under Different Attack Scenarios.}
\label{tab:far_attacks}
\setlength{\tabcolsep}{8pt}
\renewcommand{\arraystretch}{1.5}
\begin{tabular}{lccc}
\hline
\textbf{Method} &
\textbf{\makecell{Mimicry \\Attack}} &
\textbf{\makecell{Artificial Replication \\Attack}} &
\textbf{\makecell{Puppet \\Attack}} \\
\hline
Live20R~\cite{zkteco_live20r}  
& N / A 
& 15.73\% 
& 100\% \\
% Capacitive-Based 
% & -- 
% & -- 
% & -- \\
\textbf{BioMoTouch} 
& \textbf{0.51\%} 
& \textbf{0.90\%} 
& \textbf{0.51\%} \\
\hline
\end{tabular}
\end{table}

\begin{figure}[t]
  \centering
  \includegraphics[width=0.7\linewidth]{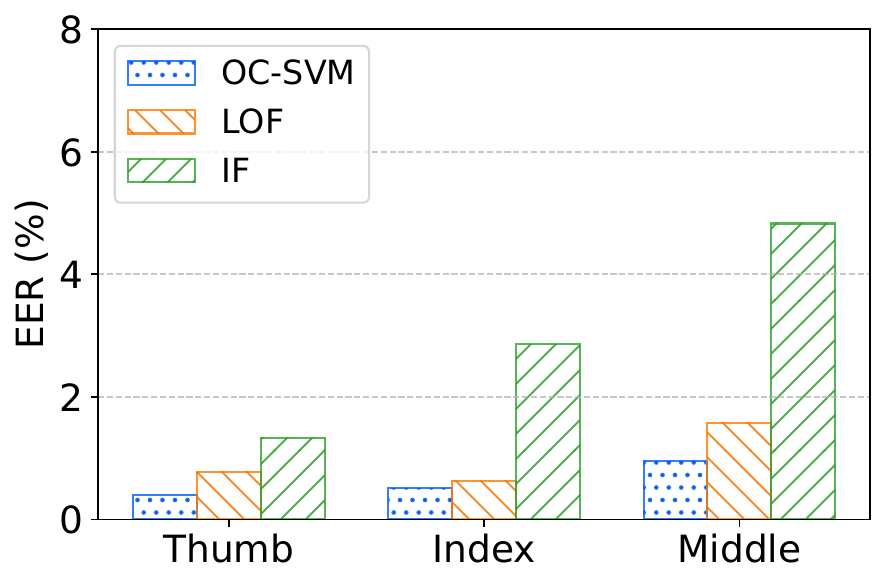} 
  %\vspace{-4pt}
  \caption{EERs of different one-class classifiers across fingers.} 
  \label{fig:finger} 
\end{figure}

\subsubsection{Performance Over Time}
Using \textit{Dataset-6}, we evaluate the temporal stability of BioMoTouch over time. As described above, press interactions are recollected from 8 subjects at five time points (Weeks 1-5) following the initial enrollment, with consistent settings and user posture across all sessions. Fig.~\ref{fig:overtime} reports the authentication performance measured in terms of EER.
Our method maintains consistently low error rates throughout five weeks, with EER values of 0.27\%, 0.53\%, 1.08\%, 0.80\%, and 0.98\%, respectively. In contrast, the capacitive-screen-based variant exhibits markedly higher EERs across all weeks (1.07\%-5.59\%) and shows more pronounced temporal variation. 
These results suggest that BioMoTouch is less affected by temporal variations introduced by repeated usage over weeks, whereas the capacitive-only variant is more sensitive to such changes.

\subsubsection{Impact of Different Fingers}
To evaluate the effect of finger selection, we independently train and evaluate a user-specific authenticator for each finger using \textit{Dataset-7}, and report EERs obtained with different one-class classifiers across three fingers. 
As shown in Fig.~\ref{fig:finger}, OC-SVM achieves the lowest EERs, ranging from 0.40\% to 0.95\%, followed by LOF with EERs between 0.63\% and 1.58\%, while IF yields substantially higher error rates from 1.33\% to 4.83\%.
Notably, the middle finger consistently exhibits the highest EERs across all classifiers, reaching 0.95\% (OC-SVM), 1.58\% (LOF), and 4.83\% (IF). This degradation is likely because most participants are not accustomed to authenticating with the middle finger, leading to less natural interactions and reduced consistency in the extracted behavioral features. Despite this effect, the system remains functional across different finger choices.

\begin{figure*}[t!]
    \begin{minipage}[t]{0.32\linewidth}
        \centering
        \includegraphics[width=\textwidth]{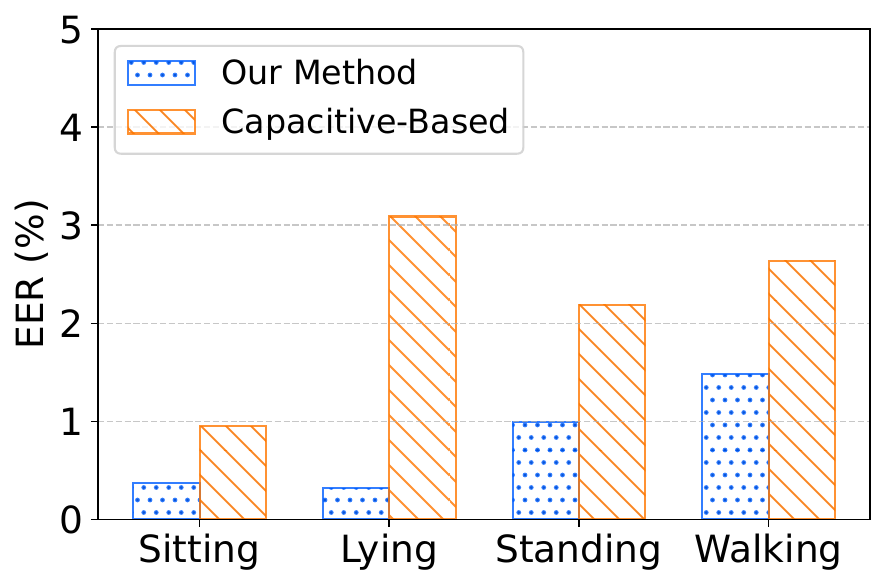} 
        \vspace{-2pt}
        \caption{EERs under different user postures.\\} 
        \label{fig:posture} 
    \end{minipage}
    \hfill   
    \begin{minipage}[t]{0.32\linewidth}
        \centering
        \includegraphics[width=\textwidth]{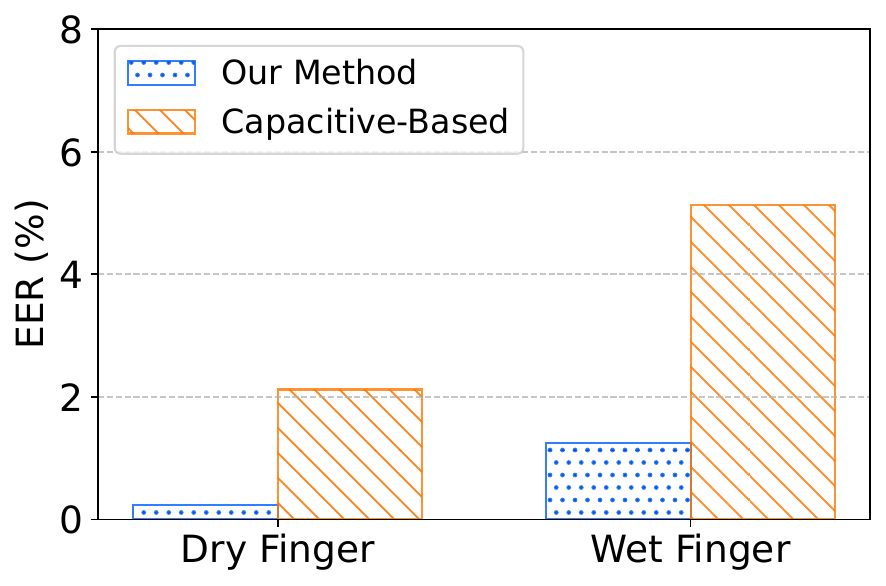}
        \vspace{-2pt}
        \caption{EERs under dry and wet finger conditions.}
        \label{fig:dry_wet}
    \end{minipage}
    \hfill
    \begin{minipage}[t]{0.32\linewidth}
        \centering
        \includegraphics[width=\textwidth]{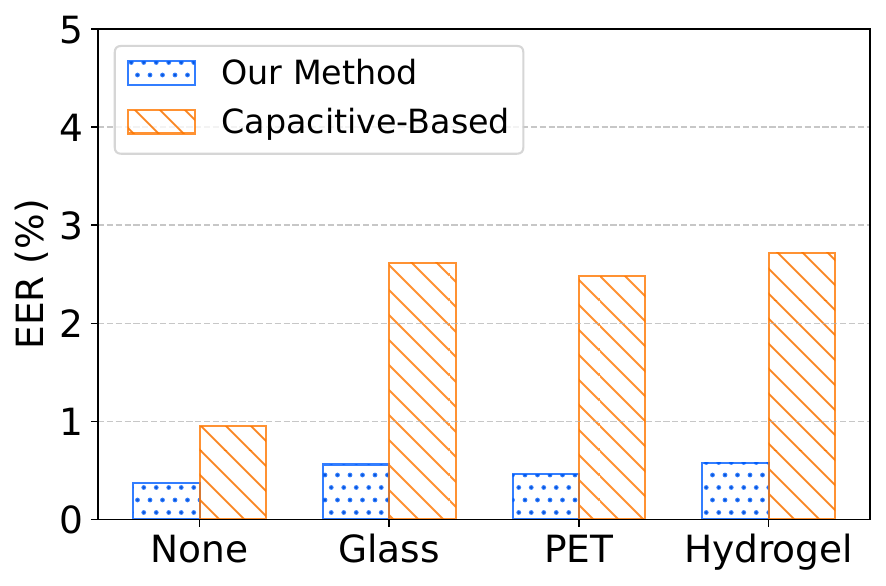}
        \vspace{-2pt}
        \caption{EERs under different screen protector conditions.}
        \label{fig:screen}
    \end{minipage}
    \vspace{-2pt}
\end{figure*}

\subsubsection{Impact of Different Posture}
To evaluate robustness against posture variations in realistic scenarios, we train and test a user-specific authenticator using \textit{Dataset-8}.
We compare the authentication performance of BioMoTouch with a capacitive-screen-based method under different user postures, including sitting, lying, standing, and walking. As shown in Fig.~\ref{fig:posture}, BioMoTouch consistently outperforms the capacitive-based method across all postures. Specifically, BioMoTouch achieves EERs of 0.37\% (Sitting), 0.32\% (Lying), 0.99\% (Standing), and 1.48\% (Walking), whereas the capacitive-screen-based method yields higher error rates of 0.95\%, 3.09\%, 2.19\%, and 2.64\%, respectively.
The lying posture yields higher EERs especially for capacitive-based methods, likely due to reduced and less stable touch pressure in a fully reclined position, which increases the mismatch between enrollment and testing features. Walking also degrades performance for both approaches because of increased body motion and device instability. Despite these effects, BioMoTouch is less sensitive to posture-induced variations.

\subsubsection{Impact of Finger Moisture}
Using \textit{Dataset-9}, we evaluate the impact of finger moisture conditions on authentication performance by comparing BioMoTouch with a capacitive-screen-based method under dry and wet finger settings. As shown in Fig.~\ref{fig:dry_wet}, BioMoTouch maintains relatively stable performance across moisture conditions, whereas the capacitive-based method is more sensitive to finger wetness.
Specifically, BioMoTouch achieves EERs of 0.27\% (dry) and 1.24\% (wet), while the capacitive-screen-based method exhibits higher error rates of 2.12\% and 5.13\%, respectively. The performance degradation under wet conditions is more pronounced for the capacitive-based method, likely due to moisture-induced variations in capacitive sensing, whereas BioMoTouch shows improved robustness to such variations.

\subsubsection{Impact of Screen Protectors}
We use \textit{Dataset-10} to evaluate the impact of screen protectors on authentication performance by comparing BioMoTouch with a capacitive-based method under four conditions: no protector (None), tempered glass (Glass), PET film (PET), and hydrogel film (Hydrogel). 
As shown in Fig.~\ref{fig:screen}, BioMoTouch maintains consistently low EERs across all screen protector types, indicating minimal sensitivity to the presence of protective layers, whereas the capacitive-screen-based method exhibits noticeably higher sensitivity.
Specifically, BioMoTouch achieves EERs of 0.37\% (None), 0.56\% (Glass), 0.46\% (PET), and 0.58\% (Hydrogel). In contrast, the capacitive-screen-based method yields substantially higher error rates of 0.95\%, 2.62\%, 2.48\%, and 2.72\%, respectively. The observed performance degradation of the capacitive-based method is likely attributable to changes in capacitive coupling and signal attenuation introduced by different protective layers. By comparison, BioMoTouch is less affected by such variations, demonstrating improved robustness to screen protector interference.

\section{Conclusion}

In this paper, we present BioMoTouch, a multi-modal touch-based behavioral authentication framework that captures fine-grained touch interaction dynamics by jointly modeling capacitive touchscreen signals and inertial measurements. Operating implicitly during natural interactions and requiring no additional hardware, BioMoTouch can be deployed on commodity mobile devices and supports both standalone authentication and seamless integration as an auxiliary signal within existing touch-based authentication systems.
Extensive evaluations demonstrate that BioMoTouch achieves strong authentication performance, with a balanced accuracy of 99.71\% and an equal error rate of 0.27\%. Moreover, BioMoTouch exhibits robust resistance to advanced adversarial behaviors, maintaining false acceptance rates below 0.90\% under artificial replication, mimicry, and puppet attacks. These results indicate that multi-modal modeling of touch interaction behaviors provides an effective and practical direction for strengthening mobile authentication systems against real-world adversarial threats.

\bibliographystyle{IEEEtran}
\bibliography{references/ref}

\end{document}